\documentclass[10pt,onecolumn]{article}
\usepackage{color}
\usepackage{hyperref}
\usepackage{tablefootnote}
\usepackage{subfigure}
\usepackage{epstopdf}
\usepackage{multirow,graphicx,array}
\usepackage{datetime}
\usepackage{lmodern}
\usepackage{tabularx}
\usepackage{xcolor}
\usepackage{listings}
\usepackage{pdfpages}
\usepackage{amssymb}
\usepackage{url}
\usepackage{gensymb}
\usepackage{tikz}
\usepackage{amssymb}
\usepackage{pifont}
\usepackage{amsmath}
\usepackage[T1]{fontenc}
\usepackage[utf8]{inputenc}
\usepackage{authblk}

\title{TouchSignatures: Identification of User Touch Actions and PINs Based on Mobile Sensor Data via JavaScript}
\author{Maryam Mehrnezhad, Ehsan Toreini, \\Siamak F. Shahandashti, Feng Hao\\\small\{{m.mehrnezhad, ehsan.toreini, siamak.shahandashti, feng.hao\}@ncl.ac.uk}} 
\affil{School of Computing Science, Newcastle University, Newcastle upon Tyne, UK}
\date{}

\begin{document}
\maketitle
%==========================================
%==========================================
\begin{abstract}
Conforming to W3C specifications, mobile web browsers allow JavaScript code in a web page to access \textit{motion and orientation} sensor data without the user's permission. The associated risks to user security and privacy are however not considered in W3C specifications. In this work, for the first time, we show how user security can be compromised using these sensor data via browser, despite that the data rate is 3 to 5 times slower than what is available in app. We examine multiple popular browsers on Android and iOS platforms and study their policies in granting permissions to JavaScript code with respect to access to motion and orientation sensor data. Based on our observations, we identify multiple vulnerabilities, and propose \emph{TouchSignatures} which implements an attack where malicious JavaScript code on an attack tab listens to such sensor data measurements. 
Based on these streams, \emph{TouchSignatures} is able to distinguish the user's touch actions (i.e., tap, scroll, hold, and zoom) and her PINs, allowing a remote website to learn the client-side user activities.  
We demonstrate the practicality of this attack by collecting data from real users and reporting high success rates using our proof-of-concept implementations. 
We also present a set of potential solutions to address the vulnerabilities. The W3C community and major mobile browser vendors including Mozilla, Google, Apple and Opera have acknowledge our work and are implementing some of our proposed countermeasures.  

%==========================================
\end{abstract}
%\end{frontmatter}
%==========================================
\section{Introduction}
\label{intro}
\label{AccesstoMobileSensors}
%==========================================
\begin{table}[t]
\centering
\begin{tabular}{llll}
\hline
Work & Sensor & Identification item & Access \\
\hline
PIN Skimmer \cite{PINCamera} & Camera, Mic& PINs & in-app\\
PIN Skimming \cite{SkimLight} & Light & PINs & in-app\\
Keylogging by Mic \cite{KeyMic} & Mic& Keyboard, PINs & in-app\\
ACCessory \cite{accessory} & Acc & Keyboard, Area taps& in-app\\
Tapprints \cite{Tapprints} & Acc, Gyr& Keyboard, Icon taps& in-app\\
Acc side channel \cite{Tapprints2}& Acc& PINs, Patterns& in-app\\
Motion side channel \cite{Motionattack} & Acc, Gyr & Keyboard, PINs &in-app\\
TapLogger \cite{taplogger}& Acc, Ori& PINs& in-app\\
TouchLogger \cite{touchlogger} & Ori &  PINs& in-app\\
\hline
\emph{TouchSignatures} & Motion, Ori& Touch actions, PINs& in-browser\\
\hline
\end{tabular}
\caption {Brief description of \emph{TouchSignatures} and in-app sensor-based Password/ PIN identifiers. 
Acc = accelerometer, Gyr = gyroscope, and Ori = Orientation. 
Motion streams are a set of measurements which are accessible within browsers and include accelerometer, accelerometer-including-gravity, and rotation rate (see Section \ref{sensordetail}). }
\label{table:comp}
\end{table}
%==========================================
\subsection{Access to mobile sensors within app}
Sensor-rich mobile devices are becoming ubiquitous. 
Smart phones and tablets are increasingly equipped with a multitude of sensors such as GPS, gyroscope, compass, and accelerometer. 
Data provided by such sensors, combined with the growing computation capabilities of modern mobile devices enable richer, more personalised, and more usable apps on such devices. 
On the other hand, access to the sensor streams provides an app running in the background a side channel.  
Listening to mobile sensor data via a background process either for improving user security \cite{Tap-Wave-Rub,TouchMe,SilentSense,Progressiveauthentication:,Audio:Light,All:Sensors,mobileiden,useriden} or attacking it \cite{touchlogger,taplogger,Tapprints,accessory,Tapprints,Speech:Gyr} has been always interesting for researchers. 

Listening to the sensor data through a malicious background process may enable the app to compromise the user security.
Here, we present Table \ref{table:comp} and briefly describe the existing in-app sensor-based password/PIN identifiers. Some of the existing works in Table \ref{table:comp} try to identify PINs and Passwords by using sensors such as light, camera and microphone \cite{SkimLight,PINCamera,KeyMic}. 
In this paper, we are interested in the use of accelerometer and gyroscope sensors as a side channel to learn about users PINs and Passwords \cite{Tapprints,accessory,Tapprints2,touchlogger,taplogger}. 
%==========================================
\subsection{Access to mobile sensors within browser}
All these attacks suggest to obtain sensor data through a background process activated by a mobile app, which requires installation and user permission. By contrast, \emph{TouchSignatures} suggests to record the sensor measurements via JavaScript code without any user permission. This is the first report of such a JavaScript-based attack. This attack is potentially more dangerous than previous app-based attacks as it does not need any user permission for installation to run the attack code. 

Mobile web applications are increasingly provided access to more mobile resources, particularly sensor data. Client-side scripting languages such as JavaScript are progressively providing richer APIs to access mobile sensor data. Currently, mobile web applications have access to the following sensor data: geolocation~\cite{W3CGPS}, multimedia (video cameras, microphones, webcams)~\cite{W3CCamera}, light~\cite{W3CLight}, and device motion and orientation~\cite{W3CMotion}. 

W3C specifications discuss security and privacy issues for some mobile sensors such as GPS and light. For example, 
the working draft on ambient light events explicitly discuss security and privacy considerations as follows~\cite{W3CLight}: ``The event defined in this specification is only fired in the top-level browsing context to avoid the privacy risk of sharing the information defined in this specification with contexts unfamiliar to the user. For example, a mobile device will only fire the event on the active tab, and not on the background tabs or within iframes''. 
The geolocation API on the other hand, requires explicit user permission to grant access to the web app due to security and privacy considerations. 

On the other hand, security and privacy issues regarding motion and orientation sensor data have not been as readily evident to the W3C community and browser vendors as those of the sensors discussed above. 
Interestingly, in contrast to geolocation and ambient light sensors, there is no \textit{security and privacy considerations} section in the W3C working draft on motion and orientation sensors~\cite{W3CMotion}.
JavaScript code in a web page is given full access to motion and orientation sensor streams on mobile devices without needing to ask for user permission. This opens the door for attackers to compromise user security by listening to the motion and orientation sensor data as we present in this paper. 
%==========================================
\subsection{Access to mobile sensors within app vs. browser}
\label{VS}
The in-browser sensor data access that the W3C specification allows is heavily restricted in multiple ways. 
First, the access is restricted to only two types of streams: the \emph{device orientation} which supplies the physical orientation of the device, and the \emph{device motion} which represents the acceleration of the device. 
Motion data includes sequences from accelerometer, accelerometer-including-gravity, and rotation rate \cite{W3CMotion}.  
The orientation sensor, on the other hand, derives its data by processing the raw sensor data from the accelerometer and the geomagnetic field sensor\footnote{\texttt{http://developer.android.com/guide/topics/sensors/sensors\_position.html{\allowbreak}\#sensors-pos-orient}}.

More importantly, access is also restricted to \emph{low-rate} streams which provide data with slower frequencies as compared to those provided in-app.
Here, we present two tables (Tables \ref{tbl:freq1} and \ref{tbl:freq2}) on sampling frequencies on different platforms and popular browsers.
%==========================================
\begin{table}[t]
	\centering 
	\begin{tabular}{|l| c |c|}
	\hline 
	Device/mOS & Accelerometer & Gyroscope \\
	 & Freq. (Hz) & Freq. (Hz)\\
	\hline 
	Nexus 5/Android 5.0.1& 200 & 200\\
	iPhone 5/iOS 8.2& 100 & 100\\	
	\hline
	\end{tabular}
				\caption{Maximum in-app sampling frequencies on different mobile OSs}
				\label{tbl:freq1}
\end{table}
%==========================================
The in-app frequency rates in Table \ref{tbl:freq1} for Android are obtained from running an open source program (\textit{MPLSensor.cpp} file) available on Android Git repository\footnote{\texttt{https://android.googlesource.com/platform/hardware/invensense/+/{\allowbreak}android-5.0.1\_r4}}. 
And the in-app frequency rates for iOS are from \textit{system.setAccelerometerInterval()} and \textit{system.setGyroscopeInterval()} functions available on Coronalabs\footnote{\texttt{https://docs.coronalabs.com/api/library/system}}. 
For obtaining the in-browser accelerometer and gyroscope sampling rates presented in Table \ref{tbl:freq2}, we implemented our own JavaScript code (see \ref{appb}). We observed the amount of data recordable during a second in different mobile operating systems (mobile OS) and browsers. 

As it can bee seen in Table \ref{tbl:freq1}, iOS and Android limit the mentioned sensors' maximum sampling rates to 100 Hz and 200 Hz, respectively. However, the hardware is capable to sample the sensor signals at much higher frequencies (up to thousands of Hz) \cite{Speech:Gyr}. 
This reduction is to save power consumption. 
Moreover according to the results of our tests in Table \ref{tbl:freq2}, we found out that all currently available versions on different mobile browsers reduce the sampling rate even further - 3 to 5 times lower, regardless of the engine (Webkit, Blink, Gecko, etc.) that they use. Our observations on the sampling rates of different mobile browsers are mostly consistent with the results reported in \cite{Speech:Gyr}.

%==========================================
\begin{table}[t] 
	\centering 
	\begin{tabular}{|l l l| c| c|}   
		\hline
	 Device & OS & Browser & Motion     & Orientation \\
	        &    &         & Freq. (Hz) & Freq. (Hz)\\

		\hline
				\multirow{5}{*}{\begin{tabular}{c} \rotatebox[origin=c]{90}{Nexus 5} \end{tabular}} & 
                \multirow{5}{*}{\begin{tabular}{c} \rotatebox[origin=c]{90}{Android 5.0.1} \end{tabular}} & 
                Chrome  & 60 & 44\\
     			& & Opera  & 60 & 52\\
     			& & Firefox  & 50 & 50\\
     			& & Dolphin  & NA & 151\\
     			& & UC Browser & NA & 15\\
\hline
	\multirow{4}{*}{\begin{tabular}{c} \rotatebox[origin=c]{90}{iPhone 5} 
	\end{tabular}} & 
	\multirow{4}{*}{\begin{tabular}{c} \rotatebox[origin=c]{90}{iOS 8.2} 
	\end{tabular}} & 
				Safari  & 20 & 20\\
     			& & Chrome &  20 & 20\\
     			& & Dolphin  & 20 & 20\\
     			& & UC Browser & 20 & 20\\
\hline
	\end{tabular}
		\caption{Maximum in-browser sampling frequencies on different mobile OSs and browsers}
		\label{tbl:freq2}
\end{table}
%==========================================
The tight restrictions for in-browser access on sensor-related data streams seem to be put in place as a measure to strike a balance between providing too little data to be useful on one hand and too much data which can potentially compromise user security on the other hand. 
Indeed, the low-rate and processed device orientation and motion data streams provided in-browser give the impression of being the minimum needed to make applications such as game control possible in-browser, and might project a sense of security in using such in-browser access to sensor-related data in practice. 
However, in this work, for the first time, we show how user security can be compromised using device motion and orientation data provided in-browser as a side channel. 
We demonstrate how an inactive or even a minimised web page, using JavaScript, is able to listen to and silently report the device motion and orientation data about a user who is working on a separate tab or a separate app on the device. 
Moreover, we show that the reported data, although restricted in multiple ways as discussed before, is sufficient to recognise the user's \emph{touch actions} such as tapping, holding, scrolling (up, down, left, and right), and zooming (in and out), and eventually the user's PINs on the separate tab/app. 

Note that neither Android nor iOS explicitly requires user permission to access such sensor data at the time when the browser is installed. 
Furthermore, none of the browsers seek user permission or even notify the user when such sensor data is provided to a JavaScript-enabled web page. 
Consequently, the user is completely oblivious to such an attack that compromises her security. 
At the same time, users increasingly use web browsers on their mobile devices to access services such as online banking and healthcare services which involve personal and highly sensitive information. 
These facts demonstrate the potential damage that may be caused by attacks such as ours and stress the urgent need for major mobile operating systems and browser developers and also W3C standards to address this problem. 
%==========================================
\subsection{Contributions}
In this work,  we initiate the first study on the possibility of attacks compromising user security via web content, and demonstrate weaknesses in W3C standards, and also mobile OS and browser policies which leave the door open for such exploits. 
In particular, the  main contributions of this work are as follows: 
\begin{itemize}
	\item We examine multiple popular browsers on both Android and iOS platforms and study 1) their sampling frequencies, and 2) their policies in granting permissions to JavaScript code with respect to access to orientation and motion sensor data. 
	Based on these examinations, we identify multiple vulnerabilities 
	which could be potentially exploited in different attack scenarios. 
	
	\item Based on our findings, we propose \emph{TouchSignatures} which includes attacks that compromise user security through malicious JavaScript code by listening to orientation and motion sensor data streams. 
	Our attack is designed in two phases: 
	1) identifying user's touch actions (e.g.\ tap, scroll, hold, and zoom), and
	2) identifying user's PINs. 
	We demonstrate the practicality of the above two-phase attack by collecting data from real users and reporting high success rates using our proof-of-concept implementations. 
\end{itemize}
%==========================================
\newcommand{\jsadefbad}{\textit{yes}} %definitely bad
\newcommand{\jsadefbadp}{\underline{\textit{yes}}} %definitely bad but partial
\newcommand{\jsaprobad}{yes} %probabbly bad 
\newcommand{\jsagood}{---}
\newcommand{\jsafamily}{$\dagger$}
%==========================================
\begin{table*}[!t] 
	\centering 
	\begin{tabular}{| c l | c c c | c c | c c |}   
		\hline 
		\multicolumn{2}{|c|}{\multirow{2}{*}{Device/mOS/Browser}} & \multicolumn{3}{c|}{Active} & 
		\multicolumn{2}{c|}{Background} &\multicolumn{2}{c|}{Locked}\\ 
		\cline{3-9}
		\multicolumn{2}{|c|}{} & same & iframe & other & same & other & same & other \\ 
		\cline{1-9} \cline{1-9} 
		\multirow{12}{*}{\begin{tabular}{c} \rotatebox[origin=c]{90}{Nexus 5/Android 5.0.1} \end{tabular}} & 
		Chrome 		& \jsaprobad& \jsadefbad & \jsagood & \jsagood & \jsagood & \jsagood & \jsagood \\ 
		
		& Opera \jsafamily	& \jsaprobad	& \jsadefbad & \jsagood & \jsagood & \jsagood & \jsagood & \jsagood \\
		
		& Firefox 	& \jsaprobad & \jsadefbad & \jsagood & \jsagood & \jsagood & \jsagood & \jsagood \\ 
		
		& Dolphin  & \jsaprobad	& \jsadefbad & \jsagood & \jsagood & \jsagood & \jsagood & \jsagood \\ 
		
		& UC Browser \jsafamily & \jsaprobad	& \jsadefbad & \jsadefbad & \jsagood & \jsagood & \jsagood & \jsagood \\ 
		
		& Baidu & \jsaprobad	& \jsadefbad & \jsadefbad & \jsadefbad & \jsadefbad & \jsadefbad & \jsadefbad \\ 
		& CM Browser & \jsaprobad	& \jsadefbad & \jsadefbad & \jsadefbad & \jsadefbad & \jsadefbad & \jsadefbad \\ 
		& Photon  & \jsaprobad &\jsadefbad&\jsadefbad&\jsadefbad&\jsagood&\jsadefbad&\jsadefbadp \\
		& Maxthon & \jsaprobad	& \jsadefbad & \jsadefbad & \jsadefbad & \jsadefbad & \jsadefbad & \jsadefbad \\ 
		& Boat  & \jsaprobad	& \jsadefbad & \jsadefbad & \jsadefbad & \jsadefbad & \jsadefbad & \jsadefbad \\ 
		& Next  & \jsaprobad	& \jsadefbad & \jsadefbad & \jsadefbad & \jsadefbad & \jsadefbad & \jsadefbad \\ 
		& Yandex & \jsaprobad &\jsadefbad&\jsagood&\jsadefbad&\jsagood&\jsadefbad&\jsagood\\
		\hline   
		\multirow{8}{*}{\begin{tabular}{c}  
				\rotatebox[origin=c]{90}{iPhone 5/iOS 8.2} \end{tabular}} & 
		Safari 	& \jsaprobad	& \jsadefbad & \jsagood & \jsagood & \jsagood & \jsadefbadp & \jsagood \\ 
		& Chrome 	& \jsaprobad	& \jsadefbad & \jsadefbad & \jsagood & \jsagood & \jsagood & \jsagood \\ 
		& Dolphin 	&\jsaprobad & \jsadefbad & \jsadefbad & \jsagood & \jsagood & \jsagood & \jsagood \\ 
		& UC Browser & \jsaprobad	& \jsadefbad & \jsagood & \jsadefbad & \jsagood & \jsadefbad & \jsagood \\ 
		& Baidu Browser	& \jsaprobad	& \jsadefbad & \jsadefbad & \jsadefbad & \jsadefbad & \jsadefbad & \jsadefbad \\
		& Maxthon & \jsaprobad	& \jsadefbad & \jsadefbad & \jsagood & \jsagood & \jsagood & \jsagood \\ 
		& Yandex &\jsaprobad & \jsadefbad & \jsadefbad & \jsagood & \jsagood & \jsagood & \jsagood \\
		& Mercury 	&\jsaprobad & \jsadefbad & \jsadefbad & \jsagood & \jsagood & \jsagood & \jsagood \\
		\hline 
	\end{tabular}
	\caption{Mobile browser access to the orientation and motion sensor data on Android and iOS under different conditions. A \jsafamily\ indicates a family of browsers (e.g., Opera and Opera Mini are considered to be in the same Opera family). A \jsadefbad\ (in italics) indicates a possible security leakage vector. A \jsadefbadp\ (in italics and underlined) indicates a possible security leakage vector only in the case when the browser was active before the screen is locked.}
	\label{tbl:access}
\end{table*}
%==========================================
\section{Examining mobile browsers}
\label{ExaminingMobileBrowsers}
%==========================================
In this Section, we report our findings on different mobile OSs and mobile browsers policies with respect to providing access to device motion and orientation sensor data to active web content. We developed JavaScript code (see \ref{appb}) that listens to and records the above sensor data streams and carried out tests on different combinations of mobile OSs and browsers. We considered both Android and iOS, and on each mobile OS we tested the major browsers plus those that are highly popular (see \ref{appa}). 
The details of our tests and our findings are summarised in Table~\ref{tbl:access}.   

Table~\ref{tbl:access} shows the results of our tests as to whether each browser provides access to device motion and orientation sensor data in different conditions. 
The culumn(s) list the device, mobile OS (mOS), and browser combination under which the test has been carried out. 
In case of multiple versions of the same browser, as for Opera and Opera Mini, we list all of them as a family in one bundle since we found that they behave similarly in terms of granting access to the sensor data with which we are concerned. 
The ``yes'' indications under ``active/same'' show that all browsers provide access to the mentioned sensor data if the browser is \emph{active} and the user is working on the \emph{same} tab as the tab in which the code listening to the sensor data resides. 
This represents the situation in which there is perhaps a common understanding that the code \emph{should} have access to the sensor data. 
In all other cases, as we discuss bellow, access to the sensor data provides a possible security leakage vector through which attacks can be mounted against user security. 
In the following we give more details on these results. 

\begin{figure*}[t]
	\centering
	\includegraphics[scale=0.23]{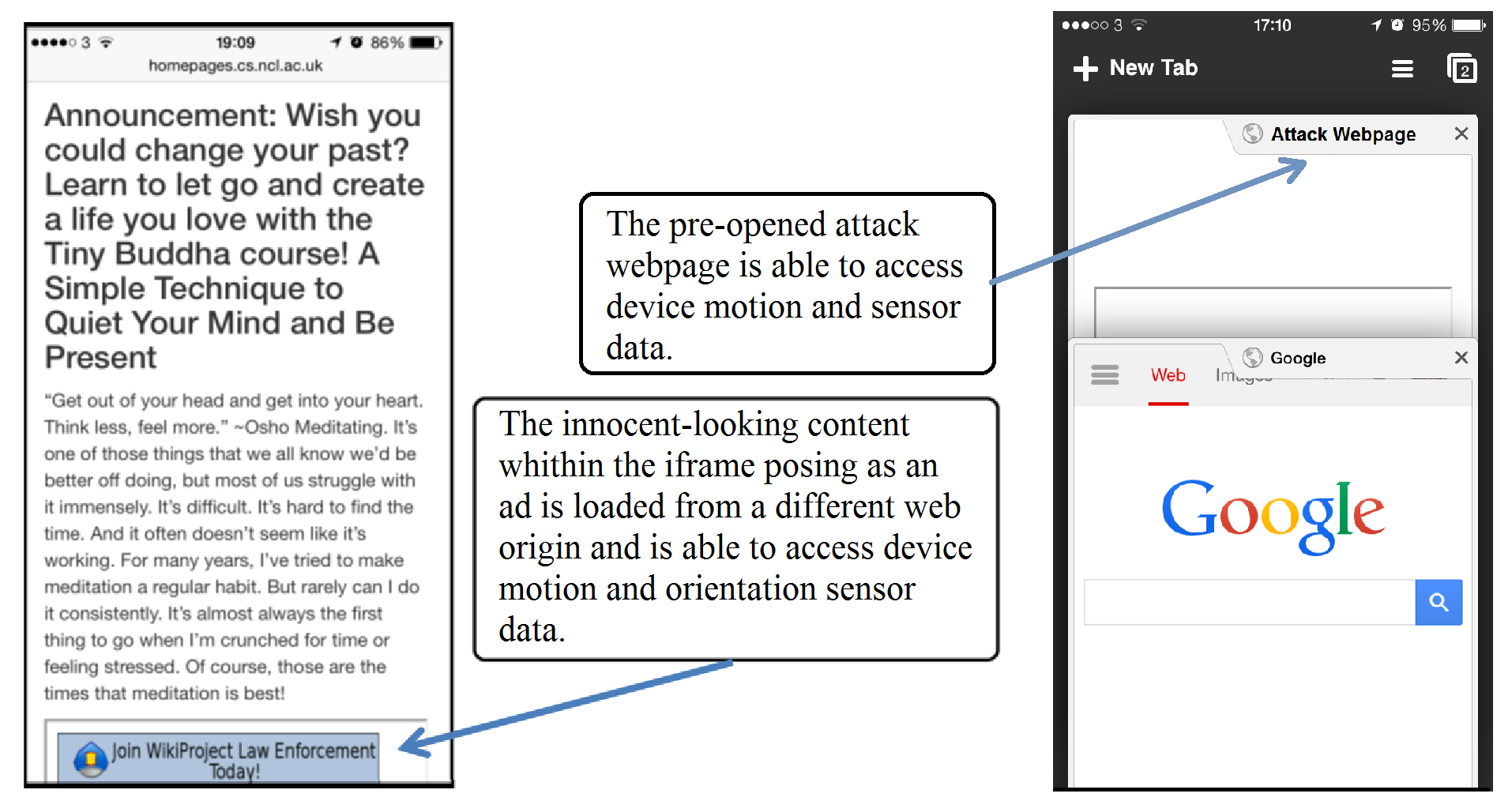}
	\caption{Left: An example of a page that includes an \texttt{iframe} (at the bottom of the page). Right: An example of a pre-opened attack page while the user is working on a different tab.
		These two examples demonstrate why \textit{iframe} and \textit{other tab} accesses can be threats to user security.}
	\label{JVA}
\end{figure*}

\paragraph{Browser-active iframe access}
HTML \emph{frames} are commonly used to divide a browser window into multiple segments, each of which can independently load a separate  web document possibly from a different web origin. 
We embedded our JavaScript listener into an HTML frame, namely \texttt{iframe}, which resided within a web page at a different web address. 
The test was to find out whether or not the listener in a separate segment of the browser window was able to access the sensor data streams if the user was interacting (via touch actions) with the content within the same tab but on a different segment of the browser window. Figure~\ref{JVA}~(left) gives an example on how an \texttt{iframe} works inside a page. The \texttt{iframe} content is loaded from a different source and is able to collect sensor data using JavaScript.
%==========================================
%==========================================
Through experiments, we found that all the browsers under test provided access to the sensor data streams in this case. 
The findings are listed in the column under ``active/iframe'' in Table~\ref{tbl:access} indicating such an access. 

\paragraph{Browser-active different-tab access}
In this test, we had the browser active and our JavaScript listener opened in a tab while the user was interacting with the content on a separate tab. Figure~\ref{JVA}~(right) gives an example of this condition. 
Interestingly, we found that in addition to most of the other browsers on Android and iOS, some major browsers such as Google Chrome on iOS provided different-tab access to the sensor data streams in this case. 
The findings are listed in the column under ``active/other'' in Table~\ref{tbl:access} indicating browser-active different-tab access. 

\paragraph{Browser-in-background access}
In this test, we first opened a web page containing our JavaScript listener and then minimised the browser.
While the browser was still running in the background, the user would interact with another app (via touch actions), or try to unlock the screen by providing a PIN or pattern input. 
We ran the test in two cases: 1) the browser had only the tab containing our JavaScript listener open, or 2) the browser had multiple tabs open including one containing our JavaScript listener. 
Surprisingly, we found that a few browsers on both the Android and iOS provided access to the sensor data streams when the user was interacting with another app. 
The findings are listed in the column under ``background'' in Table~\ref{tbl:access} indicating browser-in-background access. 

\paragraph{Screen-locked access}
In this test, we first opened a web page containing our JavaScript listener and then locked the screen.
We found that a few browsers on both Android and iOS, including Safari, provided access to the sensor data streams even when the screen was locked. 
The findings are listed in the column under ``locked'' in Table~\ref{tbl:access} indicating screen-locked access. 

We emphasise that none of the tested browsers (on Android or iOS) asked for any user permissions to access the sensor data when we installed them or while performing the experiments. 

The above findings suggest possible attack vectors through which malicious web content may gather information about user activities and hence breach user  security. 
In particular, \emph{browser-active iframe access} enables active web content embedded in HTML frames, e.g.\ posing as an advertisement banner, to discretely record the sensor data and determine how the user is interacting with other segments of the host page. 
\emph{Browser-active different-tab access} enables active web content that was opened previously and remains in an inactive tab, to eavesdrop the sensor data on how the user is interacting with the web content on other tabs. 
\emph{Browser-in-background} and \emph{screen-locked access} enable active web content that remains open in a minimised browser to eavesdrop the sensor data on how the user is interacting with other apps and on user's actions while carrying the device. 

To show the feasibility of our security attack, in the following sections, we will demonstrate that, with advanced machine learning techniques, we are able to distinguish the user's touch actions and PINs with high accuracy when the user is working with a mobile phone.
%==========================================
\section{TouchSignatures}
\label{TouchSignatures}
%==========================================
 \begin{figure}[!t]
 	\centering
    \includegraphics[scale = 0.3]{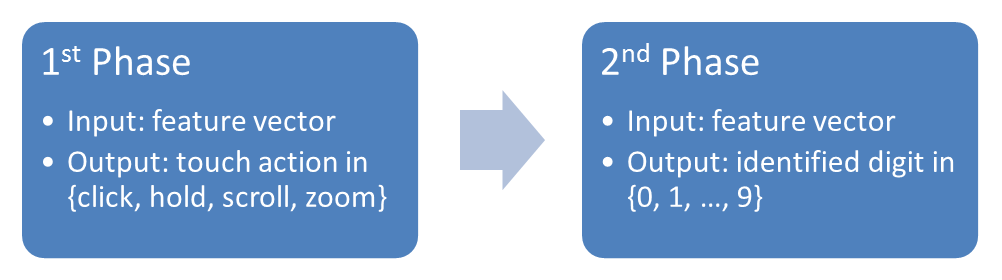}
 	\caption{\emph{TouchSignatures} overview}
 	\label{Sig}
 \end{figure}
%==========================================
\subsection{Overview}
Each user touch action, such as clicking, scrolling, and holding, and even tapping characters on the mobile soft keyboard, induces device orientation and motion traces that are potentially distinguishable from those of other touch actions. Identification of such touch actions may reveal a range of activities about user's interaction with other webpages or apps, and ultimately their PINs. 
A user's touch actions may reveal what type of web service the user is using as the patterns of user interaction are different for different web services: e.g., users tend to mostly scroll on a news web site, while they tend to mostly type on an email client. 
On known web pages, a user's touch actions might reveal which part of the page the user is more interested in. 
Combined with identifying the position of the click on a page, which is possible through different signatures produced by clicking different parts of the screen, the user's input characters could become identifiable. 
This in turn reveals what the user is typing on a page by leveraging the redundancy in human languages, or it may dramatically decrease the size of the search space to identify user passwords. 

We introduce \emph{TouchSignatures} in order to distinguish user touch actions and PINs in two phases. 
Figure \ref{Sig} shows a top level view of the two phases of \emph{TouchSignatures}. The input of \emph{TouchSignatures} system is a feature vector which we will explain later and the output is the type of the touch action (click, hold, scroll, and zoom) in phase one and the PIN digits (0 to 9) in phase two. This is the first attack in the literature that compromises user security through JavaScript access to sensor data. 

As this paper is only the first investigation on JavaScript access to sensor data on mobile devices, we limit the scope of the paper as follows. First, we only identify digital PINs rather than alphanumeric passwords. We expect it to be possible to extend our work to recognize the full alphanumeric soft keyboard, but the classification techniques will be quite different. Second, in the proof-of-concept implementation of the attack, we focus on working with active web pages, which allows us to easily identify the start and end of a touch action through the JavaScript access to the \textit{onkeydown}, and \textit{onkeyup} events. A similar approach is adopted in other works (e.g., TouchLogger \cite{touchlogger} and TapLogger \cite{taplogger}). In a general attack scenario, a more complex segmentation process is needed to identify the start and end of a touch action. This may be achieved by measuring the peak amplitudes of a signal, as done in \cite{KeyMic}. However, the segmentation process will be more complex, and we leave that to future work.
 
\subsection{In-browser sensor data detail}
\label{sensordetail}
The attack model we consider is malicious web content spying on a user via JavaScript. The web content is opened as a web page or embedded as an HTML frame in a segment of a web page. The user may be interacting with the browser or any other app given that the browser is still running in the background. We assume that the user has access to the Internet as reasonably implied by the user launching the browser app. \emph{TouchSignatures}'s client-side malicious web content collects and reports sensor data to a server which stores and processes the data to identify the user's touch actions. 
 
The sensor data streams available as per the W3C specifications \cite{W3CMotion}, i.e., device motion and orientation, as follows: 
 \begin{itemize}
 	\item device \emph{orientation} which provides the physical orientation of the device, expressed as three rotation angles: alpha, beta, and gamma, in the device's local coordinate frame,
 	
 	\item device \emph{acceleration} which provides the physical acceleration of the device, expressed in Cartesian coordinates: x, y, and z, in the device's local coordinate frame, 
 	
 	\item device \emph{acceleration-including-gravity} which is similar to acceleration except that it includes gravity as well,
 	
 	\item device \emph{rotation rate} which provides the rotation rate of the device about the local coordinate frame, expressed as three rotation angles: alpha, beta, and gamma, and
 	
 	\item \emph{interval} which provides the constant rate with which motion-related sensor readings are provided, expressed in milliseconds.
 \end{itemize}
 
The device's local coordinate frame is defined with reference to the screen in its portrait orientation: x is horizontal in the plane of the screen from left of the screen towards right; y is vertical in the plane of the screen from the bottom of the screen towards up; and z is perpendicular to the plane of the screen from inside the screen towards outside. Alpha indicates device rotation around the z axis, beta around the x axis, and gamma around the y axis, all in degrees. 
 
To design \emph{TouchSignatures}, we employ the supervised learning approach, i.e., train a machine learning system based on labelled data collected from the field. 
Consistent with the attack model discussed above, we developed a suite of applications including a client-side JavaScript program in a web page that records the sensor data and a server-side database management system (DBMS) that captures and stores user sensor data in real-time. 
Subsequently, we recruited different groups of users\footnote{All experiments in this paper were ethically approved by the Ethical Review Committee at Newcastle University, UK.} and collected sensor data samples for different touch actions and PINs, using our client-side web page that we developed for data collection purposes, while in real-time the captured sensor data was reported to and stored at our server-side database. 
Eventually, we extracted a set of descriptive features from the sensor data and trained a machine learning system for \emph{TouchSignatures} which includes multiple classifiers. 
In the following, we give the details of our application implementation, experiments, feature extraction, and training algorithms. 
\subsection{Application Implementation}
%==========================================
\begin{figure*}[t]
	\centering
	\includegraphics[scale = 0.17]{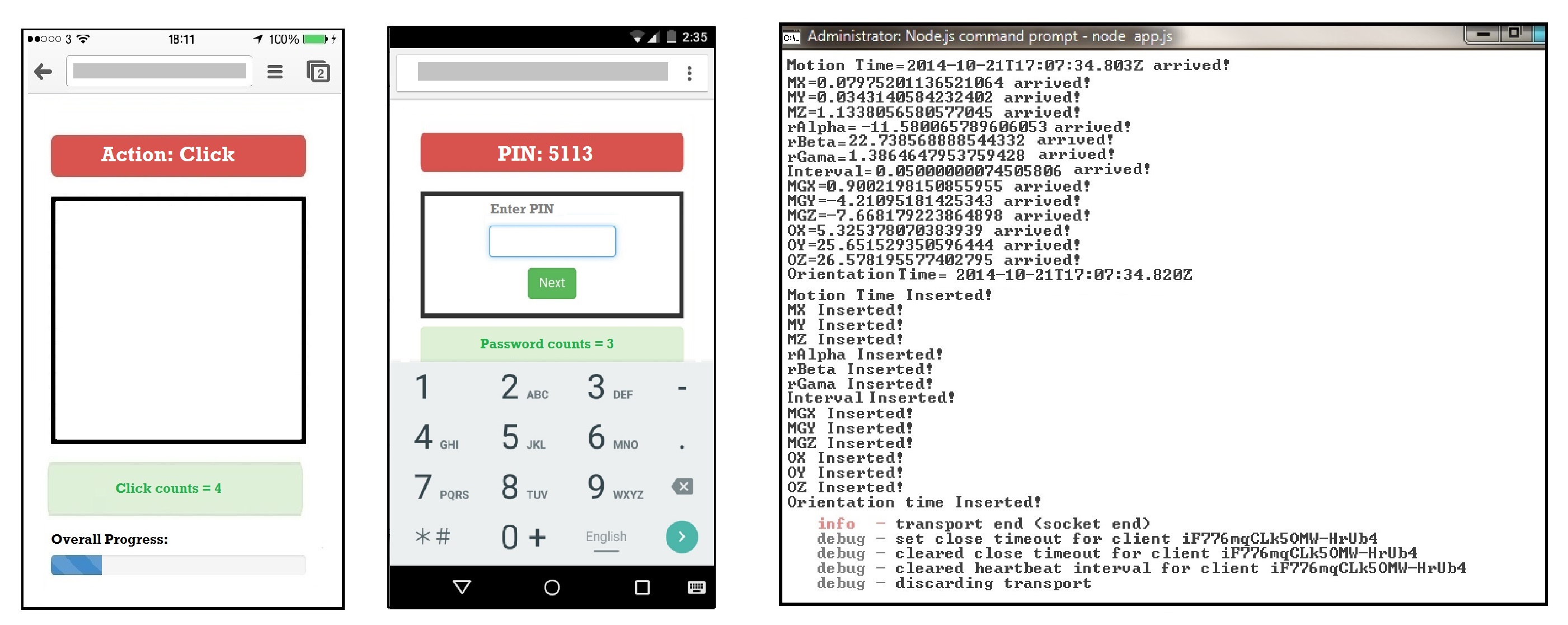}
	\caption{The client side GUIs presented to the user during data collections (left for Touch actions, and centre for PINs), and the data received at the server side (right).}
	\label{GUI}
\end{figure*}
%==========================================
\paragraph{Client side}
On the client side, we developed a \emph{listener}, which records sensor data streams, and a web page \emph{interface}, which is used to collect labelled data from the subjects in our experiment. 
The implementation is in JavaScript. 
The listener mainly includes the following components: 
an event listener which is fired on page load and establishes a socket connection between the client and server using \texttt{Socket.IO}\footnote{\texttt{www.socket.io}}, an open source JavaScript library supporting real-time bidirectional communication which runs in browser on the client side; and 
two event listeners on the window object, fired on device motion and device orientation Document Object Model (DOM) events (called \texttt{devicemotion} and \texttt{deviceorientation}), which send the raw sensor data streams to the server through the established socket.  
The sensor data streams are sent continuously until the socket is disconnected, e.g.\ when the tab that loads the listener is closed. The code is presented in \ref{appb}.

The (user) interface sits on top of the listener and is used for data collection. We developed an HTML~5 compliant page including JavaScript using \texttt{bootstrap}\footnote{\texttt{www.getbootstrap.com}} (a popular framework for web app creation). Data collection occurs in two rounds (for touch actions and PINs) and multiple steps. In each step the user is instructed to perform a single touch action or enter a 4-digit PIN. Sensor data from the touch actions and PINs are collected from the user successively. The label describing the type of the task or the digits in the PIN and timing information for the tasks is reported to the server. The GUI includes a concise instruction to the user as to what the user needs to do at each step. Snapshots of the GUIs presented to the users in the two different phases are illustrated in Figure~\ref{GUI}~(left, and centre). More details can be found in Sections \ref{Ex1} and \ref{Ex2}.

\paragraph{Server side}
On the server side, we developed a \emph{server} to host the data and handle communications, and a \emph{database} to handle the storage of the captured sensor data continuously. 
The server is implemented using \texttt{Node.js}\footnote{\texttt{www.nodejs.org}}, which is capable of supporting data intensive applications in real-time. 
The \texttt{Socket.IO} JavaScript library sits on \texttt{Node.js} and handles the communications with the client; see Figure~\ref{GUI}~(right). 
For the DBMS we have opted to implement a NoSQL database 
on MongoLab\footnote{\texttt{www.mongolab.com}}. 
NoSQL databases are document-oriented, rather than relational. They are known for being capable to handle high-speed streams of data in real-time. 
MongoLab is a cloud-based database-as-a-service NoSQL DBMS. 
%==========================================
\subsection{Feature extraction}
\label{Fez}
In this section, we discuss the features we extract to construct the feature vector which subsequently will be used as the input to the classifier. 
We consider both time domain and frequency domain features. 
The captured data include 12 sequences:
acceleration, acceleration-including-gravity, orientation, and rotation rate, with three sequences for each sensor measurement. 
Before extracting features, to cancel out the effect of the initial position and orientation of the device, we subtract the initial value in each sequence from subsequent values in the sequence. 

\paragraph{Time domain features}
In the time domain, we consider both the raw captured sequences and the (first order) derivative of each sequence. 
The rationale is that each sequence and its derivative include complementary information on the touch action.
To calculate the derivative, since we have low frequency sequences, we employ the basic method of subtracting each value from the value appearing immediately afterwards in the sequence. 
That is, if the sequence values are represented by $v_i$, the derivative sequence is defined as $d_i=v_i-v_{i-1}$. 

For the device acceleration sequences, we furthermore consider the Euclidean distance between consecutive readings as a representation of the change in device acceleration. 
This is simply calculated as the following sequence: 
\[ c_i=\sqrt{(x_i-x_{i-1})^2+(y_i-y_{i-1})^2+(z_i-z_{i-1})^2} \]  
This gives us a sequence which we call the device acceleration change sequence, or DAC sequence for short. 

First we consider basic statistical features for all sequences, their derivative, and the DAC sequence. 
These features include \emph{maximum}, \emph{minimum}, and \emph{mean (average)} of each sequence and its derivative, plus those of the DAC sequence. 
We also consider the total energy of each sequence and its derivative, plus that of the DAC sequence, calculated as the sum of the squared sequence values, i.e., $E=\sum{v_i^2}$. 
Here, in total we get 102 features for each sensor reading in the time domain. Later we will add a few more features to the input of the first phase (touch actions) in Section \ref{sec:classify}. 

\paragraph{Frequency domain features}
To distinguish between sequences with different frequency contents, we applied the Fast Fourier transform (FFT) of the sequences. 
We calculated the maximum, minimum, mean, and energy of the FFT of each sequence and consider them as our frequency domain features, i.e., a total of 48 frequency domain features. 

\subsection{Classification method}
\label{sec:classifyall}
To decide which classification method to apply to our data, we implemented various classification algorithms to assess their efficiency. 
Our test classifiers included discriminant analysis, naive Bayes, classification tree, k\-NN, and ANN. Different classifiers work better in the different phases of \emph{TouchSignatures} (touch actions and PINs). The chosen classifiers in each phase are presented in Sections  \ref{sec:classify} and \ref{sec:classify2}. In both phases, we consider a generic approach and train our algorithms with the data collected from all users. Hence, our results are user-independent.
%==========================================
\section{Phase 1: Identifying user touch actions}
\label{Phase1}
In this section we present the first phase of \emph{TouchSignatures} that is able to distinguish user touch actions given access to the device orientation and motions sensor data provided by a mobile browser. 

\subsection{Touch actions set}
We consider a set of 8 most common touch actions through which users interact with mobile devices. 
These actions include: \emph{click, scroll (up, down, right, left), zoom (in, out), and hold}. They are presented in Table~\ref{table:touch-actions} along with their corresponding descriptions. 
Our experiments show that by applying machine learning techniques these actions are recognisable from their associated sensor measurements. 
%==========================================
\begin{table*}[!h]
	\centering
	\begin{tabular}{|l@{\quad}|l|}
		\hline
		Touch Action & Description \\
		\hline
		Click & Touching an item momentarily with one finger \\
		Scroll & Touching continuously and simultaneously sliding \\
		-- up, down, right, left & \quad in the corresponding direction \\ 
		Zoom & Placing 2 fingers on the screen and sliding them \\
		-- in, out & \quad apart or toward each other, respectively \\ 
		Hold  & Touching continuously for a while with one finger \\ 
		\hline
	\end{tabular}
	\caption {The description of different touch actions users perform on the touch screen of a mobile device.}
	\label{table:touch-actions}
\end{table*} 
%==========================================
\subsection{Experiments}
\label{Ex1}
We collected touch action samples from 11 users (university staff and students) using Google Chrome on an iPhone~5.
We presented each user with a brief description of the project as well as the instruction to perform each of the 8 touch actions. 
The users were provided with the opportunity of trials before the experiment to get comfortable using the web browser on the mobile phone. 
They also could ask any question before and during the experiments.  
We asked the user to remain sitting on a chair in an office environment while performing the tasks. 
The provided GUI instructed the user to perform a single touch action in each step, collecting 5 samples for each touch action in successive steps with a three-second wait between steps. 
During the experiment, the user was notified of her progress in completing the expected tasks by the count of touch actions in an overall progress bar, as shown in Figure~\ref{GUI}~(left). 

Data were collected from each user in two settings: \emph{one-hand mode} and \emph{two-hand mode}.
In the \emph{one-hand mode}, we asked the users to hold the phone in one hand, and use the same hand's thumb for touching the screen. 
In the \emph{two-hand mode}, we asked them to use both hands to perform the touch actions. 
With these two settings, we made sure that our collected data set is a combination of different modes of phone usage. Note that zoom in/out actions can only be performed in the two-hand mode. Still, we distinguish two postures: 1) when a user holds the phone using one hand and performs zoom in/out actions by using the thumb of that hand and any finger of the other hand, and 2) when a user holds the phone using both hands and performs zoom in/out by using the thumbs of both hands. We collected data for both postures. 

We had 10 samples of each of the following actions: click, hold, scroll down, scroll up, scroll right and scroll down. Five samples were collected in the one-hand mode and 5 in the two-hand mode. In addition, we collected 10 samples for each of the following two actions: zoom in and zoom out. All 10 samples were collected in the two-hand mode, with half for each of the two postures.  Each user's output was a set of 80 samples. With 11 users, we ended up with 880 samples for our set of touch actions. 
The experiment took each user on average about 45 minutes to complete. 
Each user received a \pounds 10 Amazon voucher for their contribution to the work.
%==========================================
\subsection{Classification algorithm}
\label{sec:classify}
Before discussing the algorithms used in this phase, we add another 14 features to the \emph{TouchSignatures}' time domain features. To differentiate between touch actions with a longer ``footprint'' and those with a shorter footprint, we consider a feature which represents the length (i.e., number of readings) of each dimension of the acceleration and acceleration-including-gravity sequences that contain 70\% of the total energy of the sequence. 
To calculate this length, we first find the ``centre of energy'' of the sequence as follows: $CoE = \sum{(i\,v_i^2)} / E$, where $E$ is the total energy as calculated before. 
We then consider intervals centred at $CoE$ and find the shortest interval containing 70\% of the total energy of the sequence. 
Therefore, considering both time domain and frequency domain features from Section \ref{Fez} in addition to the new ones, \emph{TouchSignatures}' final vector for phase one has 164 features in total.                     

Our evaluations show that the $k$-nearest neighbour ($k$-NN) algorithm \cite{knn} gives the best overall identification rate for our data. 
$k$-NN is a type of lazy learning in which each object is assigned to the class to which the majority of its $k$ nearest neighbours are assigned, i.e., each feature vector is assigned to the label of the majority of the $k$ nearest training feature vectors. 
A distance function is used to decide the nearest neighbours. 
The most common distance function is the \emph{Euclidean distance}, but there are other distance functions such as the \emph{city block distance} (a.k.a.\ Manhattan or taxicab distance). 
For two given feature vectors $(f_1, f_2, \ldots, f_n)$ and $(f'_1, f'_2, \ldots, f'_n)$, the Euclidean distance is defined as $\sqrt{\sum{(f_i-f'_i)^2}}$ and the city block distance as $\sum{|f_i-f'_i|}$. 

Based on the results of our evaluations, we decide to use two classifiers in two stages. 
In the first stage, the data is fed to the first classifier which is a \emph{1-NN classifier using Euclidean distance}. 
This classifier is responsible for classification of the input data into 5 categories: click, hold, zoom in, zoom out, and scroll. 
In the second stage, if the output of the first stage is scroll, then the data is fed into the second classifier which is a \emph{1-NN classifier using city block distance}. 
This classifier is responsible for classification of a scroll into one of the 4 categories: scroll up, scroll down, scroll right, and scroll left. 
We used a 10-fold cross validation approach for all the experiments. 
%==========================================
\subsection{Results}
In this section we show the results obtained from the cross validation of the collected user data by presenting the identification rates and confusion matrices for both classifiers. 
%==========================================
\begin{table}[t]
	\centering
	\begin{tabular}{|l|rrrrr|}
		\hline
		Touch & Click & Hold & Scroll & Zoom & Zoom \\
		action&  &  & & in\ \  &  out\ \ \\
		\hline
		Click& \textbf{78.18\%} & 5.45\%& 2.73\%& 0\%& 0\%\\
		Hold& 10.90\% & \textbf{88.18\%} &  0.68\%& 1.81\%& 1.82\%\\
		Scroll& 7.27\% &  2.72\%& \textbf{95.91\%} & 0.90\%& 0.90\%\\
		Zoom in & 0\% & 1.82\%& 0.23\%& \textbf{71.82\%} &  20.90\%\\
		Zoom out & 3.64\% & 1.82\% & 0.45\% &  25.45\% & \textbf{76.36\%} \\
		\hline
		Total& 100\%& 100\%& 100\%& 100\%& 100\%\\
		\hline
	\end{tabular}
	\caption {Confusion matrix for the first classifier for different touch actions}
	\label{table:conf}
\end{table}
%==================================================
Considering all scrolls (up, down, right, and left) in one category, the overall identification rate is 87.39\%. 

Table~\ref{table:conf} shows the confusion matrix for our first classifier. 
In each cell, the matrix lists the probability that the classifier correctly labels or mislabels a sample in a category. 
The actual and classified categories are listed in the columns and rows of the table, respectively. 
As shown in Table~\ref{table:conf}, the worst results are for the pairs of Click and Hold
(10.9\% and 5.45\%), and also pairs of Zoom in and Zoom out 
(25.45\% and 20.9\%). 
This is expected since click and hold are very similar actions: and hold is basically equivalent to a long click. 
Zoom in and zoom out also require the user to perform similar gestures. 
Another significant value is the classifier's confusion between click and scroll (7.27\%, 2.73\%), which again is not surprising since scroll involves a gesture similar to a click. 
Apart from the mentioned cases, the rest of the confusion probabilities are nearly negligible.  
%==========================================
\begin{table}[!t]
	\centering
	\begin{tabular}{|l|cccc|}
		\hline
		Touch & Scroll & Scroll & Scroll & Scroll \\
		action &  down&  up&  right&  left\\
		\hline
		Scroll down& \textbf{57.27\%}& 19.09\%& 12.73\%& 4.55\%\\
		Scroll up&  26.36\% & \textbf{69.09\%}& 16.36\%&  6.36\%\\
		Scroll right& 9.09\%& 4.55\%& \textbf{48.18\%}& 17.27\%\\
		Scroll left& 7.27\%& 7.27\%& 22.73\%&  \textbf{71.82\%}\\
		\hline
		Total& 100\%& 100\%& 100\%& 100\%\\
		\hline
	\end{tabular}
	\caption {Confusion matrix for the second classifier for different scroll types}
	\label{table:conf_scroll}
\end{table}
%==========================================
 
Table \ref{table:conf_scroll} shows the identification rates and confusion matrix for our second classifier, respectively. 
Overall, our second classifier is able to correctly identify the scroll type with a success rate of 61.59\%. 
The classifier mostly mislabels the pairs (down, up), and (right, left), which is somehow expected since they involve similar gestures. 

The obtained results show that attacks on user privacy and security by eavesdropping sensor data through web content are feasible and are able to achieve accurate results. Further security risks could be imposed to the users if the attack tries to identify what character has been pressed on the touch screen. In phase 2 of \emph{TouchSignatures}, we show that it is indeed possible to succeed such an attack by identifying the digits entered for the user's PINs.
%==========================================
\section{Phase 2: Identifying user PINs}
\label{Phase2}
In this section, we present the second phase of \emph{TochSignatures} which is able to identify user PINs based on the motion and orientation sensor data provided by JavaScript code. As mentioned in Section \ref{intro}, classifying soft keyboard characters on touch screen has already been explored by other researchers based on the sensor data accessible through native apps. In this work, for the first time, we show that it is also possible to do that by using the sensor data obtained via JavaScript despite the fact that the available frequency is much lower. 

In this phase, we present the results of our suggested attack on both Android (Nexus 5) and iOS (iPhone 5) devices and we train two different classifiers (neural networks) for them. Note that JavaScript is able to obtain specific information about a mobile device -- for example the browser platform and the screen size are accessible via Navigator DOM\footnote{\texttt{www.w3schools.com/js/js\_window\_navigator.asp}} and Screen DOM\footnote{\texttt{w3schools.com/js/js\_window\_screen.asp}} objects, respectively. The obtained values for the tested devices are summarized in Table \ref{DOM}. Hence, though the experiments are performed using specific mobile devices, the results have general implications on all devices. 
%==========================================
\begin{table}[!t]
	\centering
\begin{tabular}{|l|l|l|}
\hline
Attribute & iPhone 5 & Nexus 5 \\
\hline
navigator.platform & iPhone & Linux armv7l\\
screen.width& 320 pixs& 360 pixs\\
screen.height & 568 pixs& 640 pixs\\
\hline
\end{tabular}
	\caption {The device information accessible via JavaScript.}
	\label{DOM}
\end{table}
%==========================================
\subsection{Digits set}
In this work, we consider a numerical keypad and leave the attack on the full keyboard as future work.
A numerical keyboard includes a set of 10 digits: 0, 1, 2, 3, 4, 5, 6, 7, 8, and 9, and a few more characters such as -, ., and \#, depending on the mobile OS. For example Figure \ref{GUI} (centre) shows a numerical keypad on an Android device.  
The idea is to identify the pressed digits in a PIN. 
Hence from a top view, once the first phase of \emph{TouchSignatures} distinguishes that the user is ``clicking'' digits on a soft keyboard, the second phase is started in order to obtain the entered digits. 

\subsection{Experiments}
\label{Ex2}
Similar to the first experiment, we asked a group of users (university student and staff) including 12 users to participate in our experiment in two parts. The first part was on an iPhone 5 and the second part was on a Nexus 5, both using Chrome. 
After giving a brief description about the study to the users, they were presented with a simple GUI (Figure \ref{GUI}, centre) asking them to enter 25 random 4-digit PINs on both devices. The 4-digit PINs were designed in a way that each number was exactly repeated 10 times in total. After entering each 4-digit PIN, the user could press a \textit{next} button to go to the next PIN. They also could keep track of their progress as the number of PINs they have entered so far was shown on the page.

In this experiment, we asked the users to remain sitting on a chair and hold the phone in the way that they felt comfortable. The collected data contained a mixture of \textit{one-hand mode} and \textit{two-hand mode} records. In the one-hand mode, the user pressed the digits with one of the fingers of the same hand with which they were holding the phone. In the two-hand mode, they pressed the digits with either the free hand, or both hands. 
We had 10 samples of each digit for each user. Since we had 10 digits, each user's output was a set of 100 samples for each device. With 12 users, the input of our classifiers was 1200 records for iPhone 5 and 1200 records for Nexus 5.
It took each user 2 minutes on average to complete each part of the experiment with preparation and explanations. It took each user less than 10 minutes to finish the whole experiment. 

\subsection{Classification algorithm}
\label{sec:classify2}
Among different classification methods, we observed that ANN (Artificial Neural Network) works significantly better than other classifiers on our dataset. 
A neural network system for recognition is defined by a set of input neurons (nodes) which can be activated by the information of the intended object to be classified. The input can be either raw data, or pre-processed data from the samples. In our case, we have preprocessed our samples by building a feature vector as described in Section \ref{Fez}. Therefore, as input, \emph{TouchSignatures}' ANN system receives a set of 150 features for each sample.   

A neural network can have multiple layers and a number of nodes in each layer. Once the first layer of the nodes receives the input, ANN weights and transfers the data to the next layer until it reaches the output layer which is the set of the labels in a classification problem. For better performance and to stop training before over-fitting, a common practice is to divide the samples into three sets: training, validation, and test sets. 
 
We trained a neural network with 70\% of our data, validated it with 15\% of the records and tested it with the remaining 15\% of our data set. We trained our data by using pattern recognition/classifying network with one hidden layer and 10,000 nodes. 
Pattern recognition/classifying networks normally use a scaled conjugate gradient (SCG) back-propagation algorithm for updating weight and bias values in training. SCG \cite{ANNbook} is a fast supervised learning algorithm based on conjugate directions. 
The results of the second phase of \emph{TouchSignatures} are obtained according to these settings.  

\subsection{Results}
%==========================================
\begin{table}[!t]
	\centering
	\begin{tabular}{c c c}
	\begin{tabular}{|c|c|c|c|}
		    \hline
		    \textbf{1} \tiny{(54\%)} & \textbf{2} \tiny{(64\%)}& \textbf{3} \tiny{(63\%)} & - \\
		    \hline
		    \textbf{4} \tiny{(81\%)} & \textbf{5} \tiny{(67\%)}& \textbf{6} \tiny{(73\%)} & . \\
		    \hline
		    \textbf{7} \tiny{(57\%)} & \textbf{8} \tiny{(74\%)}& \textbf{9} \tiny{(79\%)} & \tiny{X}\\
		    \hline
		    $*\#$ & \textbf{0} \tiny{(73\%)}& \tiny{English}& \tiny{$>$} \\
		    \hline
			\end{tabular}
		&  & 
	\begin{tabular}{|c|c|c|}
	    \hline
	    \textbf{1} \tiny{(70\%)} & \textbf{2} \tiny{(50\%)}& \textbf{3} \tiny{(59\%)}  \\
	    \hline
	    \textbf{4} \tiny{(70\%)} & \textbf{5} \tiny{(46\%)}& \textbf{6} \tiny{(56\%)}  \\
	    \hline
	    \textbf{7} \tiny{(53\%)} & \textbf{8} \tiny{(48\%)}& \textbf{9} \tiny{(67\%)} \\
	    \hline
	    $+*\#$ & \textbf{0} \tiny{(41\%)}& \tiny{$>$} \\
	    \hline
		\end{tabular}	
	\\
	Nexus 5 (Ave. iden. rate: 70\%)& & iPhone 5 (Ave. iden. rate: 56\%)\\
	\end{tabular}
	\caption {Identification rates of digits in Nexus 5 and iPhone 5.} 
	\label{table:rate_PIN}
\end{table}
%==========================================

Here, we present the output of the suggested ANN for Nexus~5 and iPhone~5, separately. 
Table \ref{table:rate_PIN} shows the accuracy of the ANN in classifying the digits presented in two parts for the two devices. 
The average identification rates for Nexus 5 and iPhone 5 are 70\% and 56\%, respectively. In general, the resolution of the data sequences on Android was higher than iOS. We recorded about 37 motion
 and 20 orientation measurements for a typical digit on Android, while there were only 15 for each sequence on iOS. This can explain the better performance of \emph{TouchSignatures} on Android than on iOS. It is worth mentioning that attacks on iPhone~5 actually are the ones with the lowest sampling rates that we observed in Table~\ref{tbl:freq2} (20Hz for both motion and orientation). Interestingly, even with readings on the lowest available sampling rate, the attack is still possible.  

%===================Android================
\begin{table}[!t]
	\centering
	\begin{tabular}{|c|c|c|c|}
	\hline
	%===============1=====================
	\begin{tabular}{ccc}
	 \textbf{1} \tiny{(54\%)} & \tiny{(8\%)} & \tiny{(0\%)}\\
	 \tiny{(15\%)} & \tiny{(0\%)} &  \tiny{(0\%)}\\
	 \tiny{(15\%)} & \tiny{(8\%)} & \tiny{(0\%)}\\
 	\tiny{-} & \tiny{(0\%)}&\tiny{-}\\
	\end{tabular} &
	%================2====================
	\begin{tabular}{ccc}
	\tiny{(8\%)}& \textbf{2} \tiny{(64\%)} & \tiny{(12\%)} \\
	\tiny{(0\%)}& \tiny{(12\%)} & \tiny{(0\%)}\\
	\tiny{(0\%)}& \tiny{(4\%)} & \tiny{(0\%)}\\
	\tiny{-} & \tiny{(0\%)}&\tiny{-}\\
	\end{tabular} &
	%================3====================
	\begin{tabular}{ccc}
	\tiny{(10\%)}&\tiny{(5\%)}& \textbf{3} \tiny{(63\%)}  \\
	\tiny{(0\%)}&\tiny{(0\%)}& \tiny{(5\%)} \\
	\tiny{(5\%)}&\tiny{(5\%)}& \tiny{(0\%)} \\
	\tiny{-} & \tiny{(5\%)}&\tiny{-}\\
	\end{tabular} &
	%====================================
    - \\
	%====================================
	%================4====================
	\hline
	\begin{tabular}{ccc}
	 \tiny{(0\%)}& \tiny{(0\%)}& \tiny{(6\%)}\\
	 \textbf{4} \tiny{(81\%)} & \tiny{(0\%)}&\tiny{(6\%)}\\
	 \tiny{(6\%)} & \tiny{(0\%)}&\tiny{(0\%)}\\
	 \tiny{-} & \tiny{(0\%)}&\tiny{-}\\
	\end{tabular} &
	%=================5===================
	\begin{tabular}{ccc}
	\tiny{(7\%)}& \tiny{(27\%)}& \tiny{(0\%)}\\
	\tiny{(0\%)}& \textbf{5} \tiny{(67\%)} & \tiny{(0\%)}\\
	\tiny{(0\%)}& \tiny{(0\%)}& \tiny{(0\%)}\\
	\tiny{-}& \tiny{(0\%)}& \tiny{-}\\
	\end{tabular} &
	%==================6==================
	\begin{tabular}{ccc}
    \tiny{(0\%)}	&\tiny{(0\%)}& \tiny{(7\%)}\\
	\tiny{(0\%)}	&\tiny{(7\%)}& \textbf{6} \tiny{(73\%)} \\
	\tiny{(0\%)}	&\tiny{(0\%)}& \tiny{(13\%)}\\
	\tiny{-}& \tiny{(0\%)}& \tiny{-}\\
	\end{tabular} &
	%====================================
    . \\
	%====================================
	%=================7===================
	\hline
	\begin{tabular}{ccc}
	 \tiny{(7\%)}&\tiny{(0\%)}&\tiny{(14\%)}\\
	 \tiny{(14\%)}&\tiny{(7\%)}&\tiny{(0\%)}\\
	 \textbf{7} \tiny{(57\%)} &\tiny{(0\%)}&\tiny{(0\%)} \\
	 \tiny{-} & \tiny{(0\%)}&\tiny{-}\\
	\end{tabular} &
	%==================8==================
	\begin{tabular}{ccc}
	\tiny{(0\%)}& \tiny{(0\%)}& \tiny{(0\%)} \\
	\tiny{(0\%)}& \tiny{(0\%)}& \tiny{(5\%)} \\
	\tiny{(5\%)}& \textbf{8} \tiny{(74\%)} & \tiny{(5\%)}\\
	\tiny{-}& \tiny{(11\%)}&\tiny{-}\\
	\end{tabular} &
	%==================9==================
	\begin{tabular}{ccc}
    \tiny{(0\%)}&\tiny{(0\%)}& \tiny{(0\%)} \\
	\tiny{(0\%)}&\tiny{(7\%)}& \tiny{(3\%)} \\
	\tiny{(0\%)}&\tiny{(3\%)}& \textbf{9} \tiny{(79\%)}\\
	\tiny{(-}&\tiny{(7\%)}& \tiny{-}\\
	\end{tabular} &
	%====================================
    \tiny{x}  \\
	%====================================
	%====================================
	\hline
	\begin{tabular}{ccc}
	& & \\
	& $*\#$ & \\
	&  & \\
	\end{tabular} &
	%=================0===================
	\begin{tabular}{ccc}
	\tiny{(0\%)}& \tiny{(0\%)}& \tiny{(0\%)}\\
    \tiny{(7\%)}& \tiny{(0\%)}& \tiny{(7\%)}\\
	\tiny{(0\%)}& \tiny{(7\%)}& \tiny{(7\%)}\\
	-& \textbf{0} \tiny{(73\%)} & -\\
	\end{tabular} &
	%====================================
	\begin{tabular}{ccc}
	& & \\
	& \tiny{English} & \\
	& & \\
	\end{tabular} &
	%====================================
    $>$ \\
	%====================================
	%====================================
	\hline
	\end{tabular}
	\caption {Confusion matrices in Nexus 5.} 
	\label{table:conf_PIN2}
\end{table}
%==========================================
%======================iOS=================
\begin{table}[!t]
	\centering
	\begin{tabular}{|c|c|c|}
	\hline
	%=================1===================
	\begin{tabular}{ccc}
 \textbf{1} \tiny{(70\%)} & \tiny{(0\%)} & \tiny{(4\%)}\\
	 \tiny{(4\%)} & \tiny{(0\%)} & \tiny{(0\%)}\\
	 \tiny{(4\%)} & \tiny{(17\%)} & \tiny{(0\%)}\\
	 	\tiny{-}& \tiny{(0\%)}& \tiny{-}\\
	\end{tabular} &
	%=================2===================
	\begin{tabular}{ccc}
	\tiny{(0\%)} & \textbf{2} \tiny{(50\%)} & \tiny{(0\%)}\\
	 \tiny{(17\%)} & \tiny{(11\%)} & \tiny{(5\%)}\\
	 \tiny{(5\%)} & \tiny{(11\%)} & \tiny{(0\%)}\\
	 	\tiny{-}& \tiny{(0\%)}& \tiny{-}\\
	\end{tabular} &
	%=================3===================
	\begin{tabular}{ccc}
	\tiny{(0\%)}& \tiny{(6\%)}&\textbf{3} \tiny{(59\%)}\\
	\tiny{(0\%)} & \tiny{(6\%)} & \tiny{(12\%)}\\
	\tiny{(0\%)} & \tiny{(6\%)} & \tiny{(6\%)}\\
		\tiny{-}& \tiny{(6\%)}& \tiny{-}\\
	\end{tabular} \\
	%==================4==================
	\hline
	\begin{tabular}{ccc}
 \tiny{(5\%)}& \tiny{(10\%)} &\tiny{(5\%)}\\
	 \textbf{4} \tiny{(70\%)} & \tiny{(0\%)} &\tiny{(0\%)}\\
	 \tiny{(0\%)} & \tiny{(5\%)} &\tiny{(0\%)}\\
	 	\tiny{-}& \tiny{(5\%)}& \tiny{-}\\
	\end{tabular} &
	%==================5==================
	\begin{tabular}{ccc}
	 \tiny{(8\%)} & \tiny{(8\%)} & \tiny{(0\%)}\\
 	\tiny{(15\%)} & \textbf{5} \tiny{(46\%)} & \tiny{(8\%)}\\
	 \tiny{(8\%)} & \tiny{(0\%)} & \tiny{(8\%)}\\
	 	\tiny{-}& \tiny{(0\%)}& \tiny{-}\\
	\end{tabular} &
	%=================6===================
	\begin{tabular}{ccc}
	\tiny{(0\%)}& \tiny{(6\%)}& \tiny{(0\%)}\\
	\tiny{(0\%)}& \tiny{(19\%)}&\textbf{6} \tiny{(56\%)} \\
	\tiny{(6\%)}& \tiny{(6\%)}& \tiny{(0\%)}\\
		\tiny{-}& \tiny{(6\%)}& \tiny{-}\\
	\end{tabular} \\
	%==================7==================
	\hline
	\begin{tabular}{ccc}
	 \tiny{(13\%)}& \tiny{(0\%)} &\tiny{(13\%)}\\
	 \tiny{(0\%)}& \tiny{(0\%)} &\tiny{(7\%)}\\
	 \textbf{7} \tiny{(53\%)} & \tiny{(0\%)} &\tiny{(0\%)}\\
	 \tiny{-} & \tiny{(13\%)} &\tiny{-}\\
	\end{tabular} &
	%=================8===================
	\begin{tabular}{ccc}
	 \tiny{(0\%)} & \tiny{(10\%)} & \tiny{(0\%)}\\
	 \tiny{(0\%)} & \tiny{(5\%)} & \tiny{(10\%)}\\
 	\tiny{(0\%)} & \textbf{8} \tiny{(48\%)} & \tiny{(5\%)}\\
	 \tiny{-} & \tiny{(24\%)} & \tiny{-}\\
	\end{tabular} &
	%==================9==================
	\begin{tabular}{ccc}
	\tiny{(0\%)}& \tiny{(0\%)}& \tiny{(10\%)}\\
	\tiny{(0\%)}& \tiny{(5\%)}& \tiny{(0\%)}\\
	\tiny{(10\%)}& \tiny{(5\%)}&\textbf{9} \tiny{(67\%)} \\
	\tiny{-}& \tiny{(5\%)}& \tiny{-}\\
	\end{tabular} \\
	%====================================
	\hline
	\begin{tabular}{ccc}
	& & \\
	& $+*\#$ & \\
	&  & \\
	\end{tabular} &
	%====================0================
	\begin{tabular}{ccc}
		\tiny{(0\%)}& \tiny{(6\%)}& \tiny{(0\%)}\\
	\tiny{(12}\%& \tiny{(0\%)}& \tiny{(12\%)}\\
	\tiny{(0}\%& \tiny{(15\%)}& \tiny{(12\%)}\\
	\tiny{-}& \textbf{0} \tiny{(41\%)} & \tiny{-}\\
	\end{tabular} &
	%====================================
	\begin{tabular}{ccc}
	& & \\
	& $>$ & \\
	& & \\
	\end{tabular} \\
	%====================================
	\hline
	\end{tabular}
	\caption {Confusion matrices in iPhone 5.} 
	\label{table:conf_PIN}
\end{table}
%==========================================

In Tables~\ref{table:conf_PIN2} and \ref{table:conf_PIN}, we show the identification results of each digit (bold in each cell), as well as confusion matrices on both devices. The general forms of the tables are according to Android and iOS numpads. As demonstrated, each digit is presented with all possible misclassifiable digits. 
As it can be observed, most misclassified cases are either in the same row or column, or in the neighbourhood of each expected digit. 

\begin{figure}[t]
	\centering
	\includegraphics[scale = 0.3]{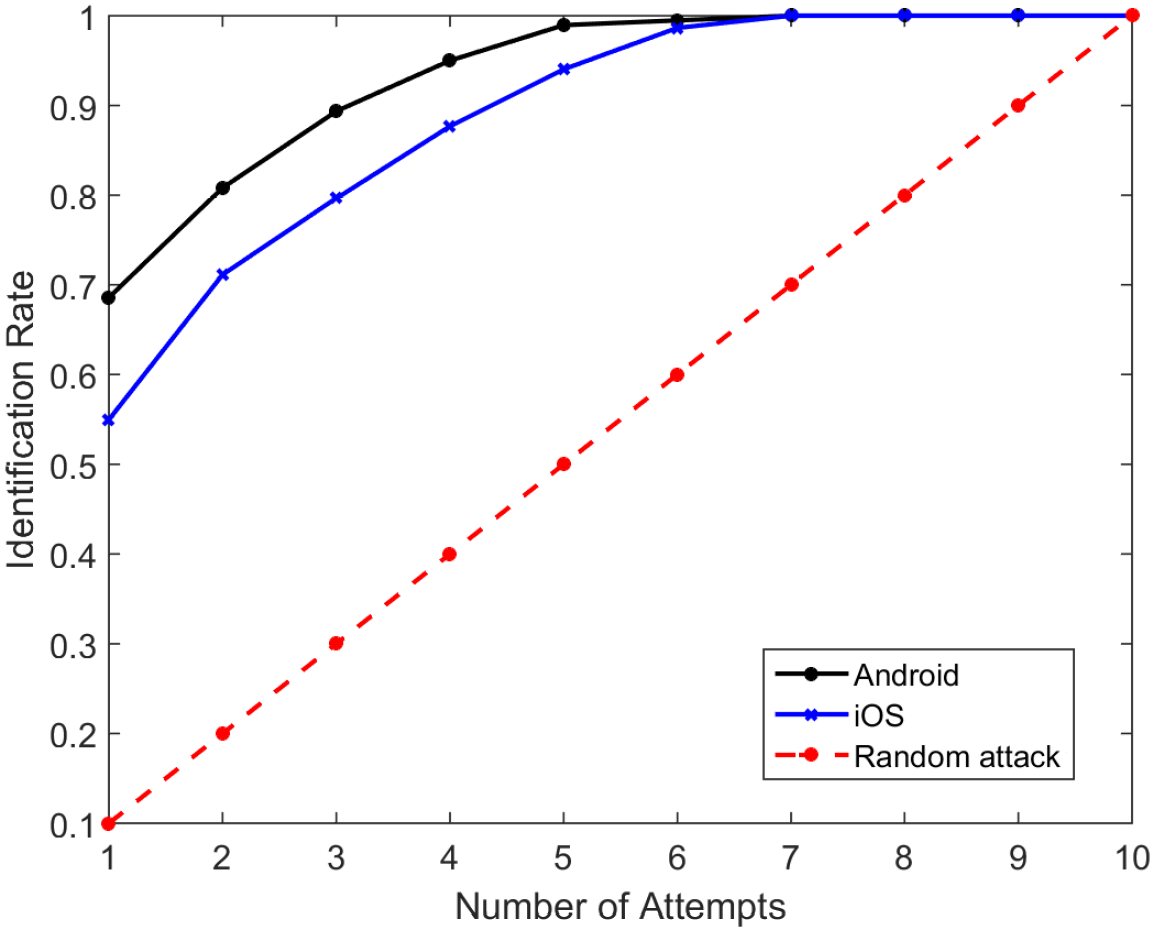}
	\caption{Average identification rate based on the number of attempts on Android and iOS vs. random guess.}
	\label{AndroidvsiOS}
\end{figure}

Note that the probability of success in finding the actual digit will significantly improve with more tries at guessing the digit. 
In fact, while the chance of the attack succeeding is relatively good on the first guess, it increases on further guesses as shown in Tables~\ref{AndroidPINGuesst} and \ref{iOSPINGuesst}.
Figure \ref{AndroidvsiOS} shows the average identification rates based on the number of guesses in Nexus~5 and iPhone~5 compared to random guessing.
As shown on the figure, \emph{TouchSignatures} can predict the correct touched digits on average in almost 90\% of the cases on Nexus~5 and 80\% of the cases on iPhone~5 in the third guess. 

%==========================================
\begin{table}[!t]
	\centering
	\begin{tabular}{|l|cccccccccc|}
	\hline
	\tiny{Digits} & 0& 1 & 2& 3& 4& 5& 6& 7& 8& 9\\
	\tiny{Attempt No.} &&&&&&&&&&\\
	\hline
	\footnotesize{First} & \footnotesize{73\%} &   \footnotesize{54\%} &  \footnotesize{64\%} &  \footnotesize{63\%} & \footnotesize{81\%}  & \footnotesize{67\%} &  \footnotesize{73\%} & \footnotesize{57\%}  & \footnotesize{74\%}  & \footnotesize{79\%}\\
	\footnotesize{Second}& \footnotesize{80\%} &   \footnotesize{69\%} &  \footnotesize{76\%} &  \footnotesize{74\%} &  \footnotesize{88\%} &  \footnotesize{93\%} &  \footnotesize{87\%} &  \footnotesize{71\%} &  \footnotesize{84\%} &  \footnotesize{86\%}\\
	\footnotesize{Third} & \footnotesize{87\%} &   \footnotesize{85\%} &  \footnotesize{88\%} &  \footnotesize{79\%} &  \footnotesize{94\%} & \footnotesize{100\%} &  \footnotesize{93\%} &  \footnotesize{86\%} &  \footnotesize{89\%} &  \footnotesize{93\%} \\
	\footnotesize{Forth} & \footnotesize{93\%} &  \footnotesize{92\%}  &  \footnotesize{96\%} &  \footnotesize{84\%} & \footnotesize{100\%} & \footnotesize{100\%} & \footnotesize{100\%} & \footnotesize{93\%}  & \footnotesize{98\%}  & \footnotesize{97\%}  \\
	\footnotesize{Fifth} & \footnotesize{100\%} &  \footnotesize{100\%}  &  \footnotesize{100\%} &  \footnotesize{89\%} & \footnotesize{100\%} & \footnotesize{100\%} & \footnotesize{100\%} & \footnotesize{100\%}  & \footnotesize{100\%}  & \footnotesize{100\%}  \\
	\footnotesize{Sixth} & \footnotesize{100\%} &  \footnotesize{100\%}  &  \footnotesize{100\%} &  \footnotesize{95\%} & \footnotesize{100\%} & \footnotesize{100\%} & \footnotesize{100\%} & \footnotesize{100\%}  & \footnotesize{100\%}  & \footnotesize{100\%}  \\
	\footnotesize{Seventh} & \footnotesize{100\%} &  \footnotesize{100\%}  &  \footnotesize{100\%} &  \footnotesize{100\%} & \footnotesize{100\%} & \footnotesize{100\%} & \footnotesize{100\%} & \footnotesize{100\%}  & \footnotesize{100\%}  & \footnotesize{100\%}  \\
	\hline
	\end{tabular}
	\caption{Identification rate based on the number of guesses that the attacker makes on Nexus~5 for each digit separately.}
	\label{AndroidPINGuesst}
\end{table}
%==========================================
\begin{table}[!t]
	\centering
	\begin{tabular}{|l|cccccccccc|}
	\hline
	\tiny{Digits} & 0& 1 & 2& 3& 4& 5& 6& 7& 8& 9\\
	\tiny{Attempt No.} &&&&&&&&&&\\
	\hline
	\footnotesize{First} & \footnotesize{41\%} &   \footnotesize{70\%} &  \footnotesize{50\%} &  \footnotesize{59\%} & \footnotesize{70\%}  & \footnotesize{46\%} &  \footnotesize{56\%} & \footnotesize{53\%}  & \footnotesize{48\%}  & \footnotesize{67\%}\\
	%============================================
	\footnotesize{Second}& \footnotesize{56\%} &   \footnotesize{87\%} &  \footnotesize{67\%} &  \footnotesize{71\%} &  \footnotesize{80\%} &  \footnotesize{62\%} &  \footnotesize{75\%} &  \footnotesize{67\%} &  \footnotesize{71\%} &  \footnotesize{76\%}\\
	%============================================
	\footnotesize{Third} & \footnotesize{69\%} &   \footnotesize{91\%} &  \footnotesize{78\%} &  \footnotesize{76\%} &  \footnotesize{85\%} & \footnotesize{69\%} &  \footnotesize{81\%} &  \footnotesize{80\%} &  \footnotesize{81\%} &  \footnotesize{86\%} \\
	%============================================
	\footnotesize{Forth} & \footnotesize{81\%} &  \footnotesize{96\%}  &  \footnotesize{89\%} &  \footnotesize{82\%} & \footnotesize{90\%} & \footnotesize{78\%} & \footnotesize{88\%} & \footnotesize{93\%}  & \footnotesize{90\%}  & \footnotesize{90\%}  \\
	%============================================
	\footnotesize{Fifth} & \footnotesize{94\%} &  \footnotesize{100\%}  &  \footnotesize{94\%} &  \footnotesize{88\%} & \footnotesize{95\%} & \footnotesize{85\%} & \footnotesize{94\%} & \footnotesize{100\%}  & \footnotesize{95\%}  & \footnotesize{95\%}  \\
	%============================================
	\footnotesize{Sixth} & \footnotesize{100\%} &  \footnotesize{100\%}  &  \footnotesize{100\%} &  \footnotesize{94\%} & \footnotesize{100\%} & \footnotesize{92\%} & \footnotesize{100\%} & \footnotesize{100\%}  & \footnotesize{100\%}  & \footnotesize{100\%}  \\
	%============================================
	\footnotesize{Seventh} & \footnotesize{100\%} &  \footnotesize{100\%}  &  \footnotesize{100\%} &  \footnotesize{100\%} & \footnotesize{100\%} & \footnotesize{100\%} & \footnotesize{100\%} & \footnotesize{100\%}  & \footnotesize{100\%}  & \footnotesize{100\%}  \\
	\hline
	\end{tabular}
	\caption{Identification rate based on the number of guesses that the attacker makes on iPhone~5 for each digit separately.}
	\label{iOSPINGuesst}
\end{table}
%==========================================
%==========================================
The high identification rates prove the feasibility of the suggested attack by \emph{TouchSignatures} and show that it is practical for a remote attacker to significantly reduce the search space for the user's PIN using JavaScript code. 

\subsection{Comparison with related works}
%==========================================
\begin{table}[t]
\centering
\begin{tabular} {|l|l|l|l|}
\hline
Work & Sensor(s) & Iden. rate & Access\\
\hline
TapLogger \cite{taplogger} & Acc, Orientation & 36.4\% & in-app \\
TouchLogger \cite{touchlogger} & Orientation & 71.5\% & in-app\\
\hline
\emph{TouchSignatures} & Motion, Orientation & 77.0\% & in-browser\\
\hline
\end{tabular}
\caption{Identification rate of phase two of \emph{TouchSignatures} (PIN) under the similar test condition as in-app attacks.}
	\label{vs}
\end{table}
%==========================================
In this section we compare the second phase of \emph{TouchSignatures}, the identification of PIN digits, with previous in-app sensor-based PIN identifiers.  
Among the works described in Table \ref{table:comp}, we choose to compare \emph{TouchSignatures} with 
TouchLogger \cite{touchlogger}, and TapLogger \cite{taplogger}, since they  use similar sensors for identifying digits on soft numerical keyboards.  

Taplogger performs its experiments on Android devices and identifies 36.4\% of the digit positions in the first attempt by using accelerometer and orientation sensors. On the other hand, TouchLogger is able to identify the digits with 71.5\% accuracy on an Android device  by using device orientation. 

TouchLogger collects around 30 samples per digit from one user, while Taplogger has the input of one user for 20 random 16-digit sequences in 60 rounds. 
However, we noticed that in these works the data has been collected from only one user. 
In general, data obtained form a single user are more consistent than those collected from a diversified group of users.  
To verify this, we performed another experiment by simulating the same test condition as described above with the Android device (Nexus 5) and asked only one user to repeat the experiment 3 times. We collected 30 samples for each digit. The results are presented in Table \ref{vs}. As expected, the identification rate of \emph{TouchSignatures} increased to 77\% in this situation, 
which is better than the results reported in TapLogger and TouchLogger.

Our results demonstrate the practicality of distinguishing the user's PIN by listening to sensor data via JavaScript code. 
Consequently, \emph{TouchSignatures} highlights the limitations of the security policies in mobile operating systems and web browsers. As a result, urgent modifications are needed in updating the security policies for granting permissions to mobile web browsers to access sensor data. 

%============================
\section{Possible solutions}
\label{Solutions}
To be able to suggest appropriate countermeasures, we need to first identify the exact entity responsible for the access control policy in each situation. 
Mobile OS access control policy decides whether the browser gets access to the device motion and orientation sensor data in the first place, no matter if the browser is active or not. 
If access is provided, then mobile browser access control policy decides whether a web app gets access to the sensor data, no matter if the web app is open in the same tab and in the same segment, in the same tab but in a different segment, or in a different tab. 
Hence any effective countermeasure must address changes in both mobile OS and browser policies with respect to access to such sensor data. 

%================================
\begin{figure}[t]
	\centering
	\includegraphics[scale=0.25]{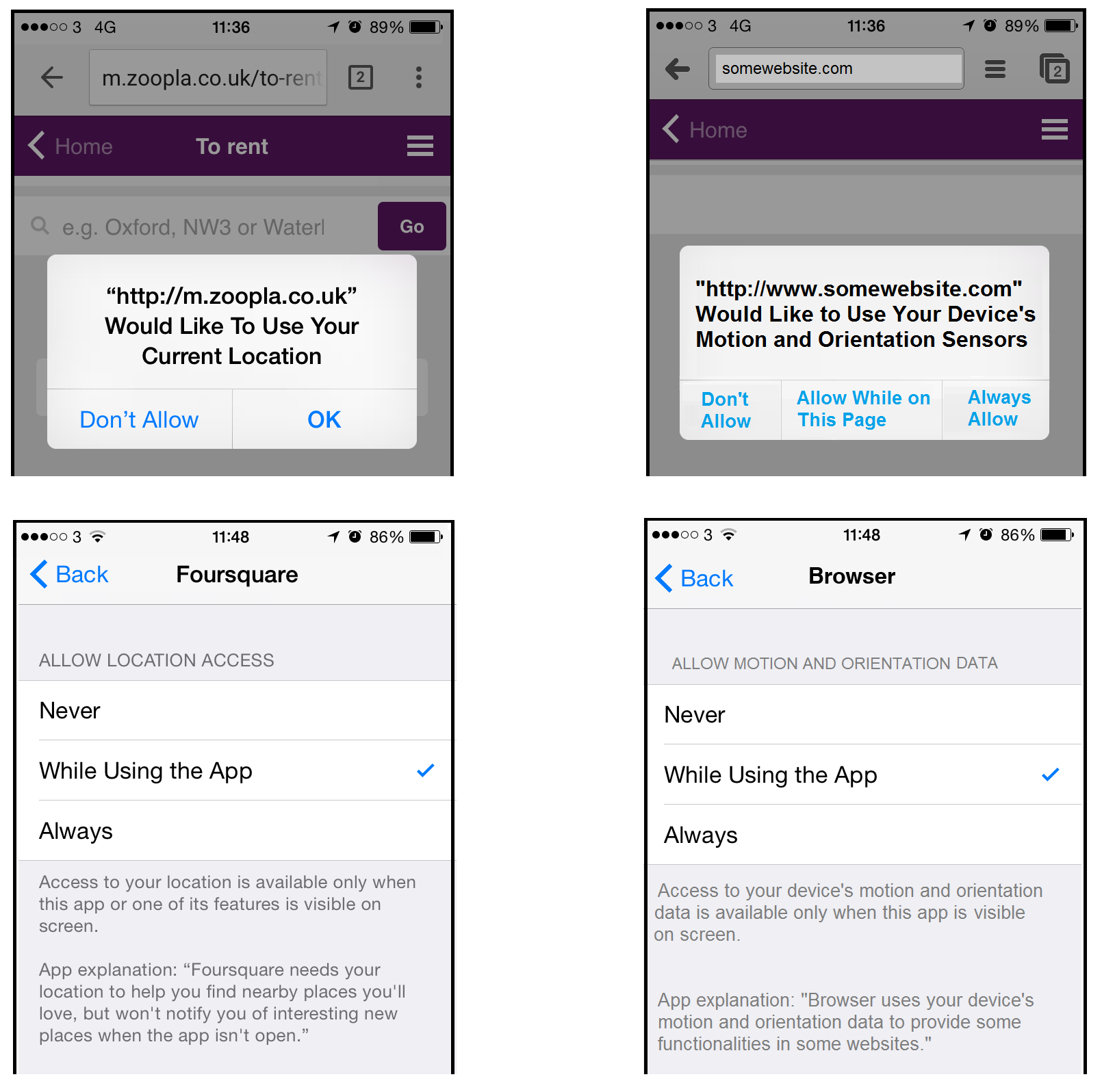}
	\caption{Left: The existing interfaces to allow the web page to access Geolocation in browser (top) and in mobile OS (down). 
	Right: Our suggested mock-up interfaces to allow web page (top) and OS setting (down) to access Motion and Orientation data in browser.}
	\label{sensoraccessF}
\end{figure}
%================================
One approach to protect user security would be to require the mobile OS to deny access to the browser altogether when the browser is not active, and require the browser to deny access to web content altogether when it is running in an inactive tab or in a segment of the page with the different web origin. 
However, this approach may be considered too restrictive as it will disallow many potential web applications such as activity monitoring for health and gaming. 

A more flexible approach would be to \emph{notify} the user when a web page is requesting access to such sensor data, and provide \emph{control} mechanisms through which the user is able to set their preferences with respect to such requests. 
This is the approach currently taken by both the mobile operating systems and browsers with respect to providing access to the device location (i.e., GPS sensor data \cite{W3CGPS}) when a web page requests such access. 
We believe similar measures for device motion and orientation would be necessary in order to achieve a suitable balance between usability and security. 
Possible (mock-up) interfaces for this countermeasure, based on existing solutions for GPS sensor data, are presented in Figure~\ref{sensoraccessF}. 
In particular, we think the user should have three options: either allow access to the browser (in the mobile OS setting) or web pages (in the browser setting) indefinitely, or allow access only when the user is working on the browser (in the mobile OS settings) or interacting with the web page (in the browser settings), or deny access indefinitely. 
These three options provided to the user seem to be neither too few to render the access control ineffective, nor too many to exhaust the user attention span. 

Furthermore, we believe raising this issue in the W3C specification would help the browsers to consider it in a more systematic and consistent way. Our suggestion for the new version of the specification is to include a section for security and privacy considerations and discuss these issues in that section properly. 
%================================
\section{Industry feedback}
We reported the results of this research to the W3C community and mobile browser vendors including Mozilla, Opera, Chromium and Apple. We discussed the identified issues with them and received positive feedback as summarized below. 

\textbf{Mozilla.} 
After we reported to Mozilla about Firefox allowing JavaScript access to sensor data within an iframe on Bugzilla, a senior platform engineer from Mozilla stated that: ``Indeed, and it should be fixed consistently across all the browsers and also the spec [W3C specification] needs to be fixed''. Subsequently, a patch has been proposed and implemented by Mozilla and is currently under test\footnote{\texttt{bugzilla.mozilla.org/show\_bug.cgi?id=1197901} (login required)}.

\textbf{Chrome \& Opera.} 
Opera uses the Chromium engine's implementation for device orientation. Therefore, fixing the problem on Opera is dependent on the fix on Chromium. We reported to both Chrome and Opera about their browsers allowing JavaScript access to sensor data within an iframe and in the other-tab. After discussing this issue on the Chromium forum, a security team member of Chrome stated that: ``It [i.e. this research] sounds like a good reason to restrict it [i.e. sensor reading] from iframes''. At the time of writing this paper, the status of our reported bug in Chromium is ``assigned''; a fix is expected to be rolled out soon. Commenting on the JavaScript access to sensor data through  \textit{other-tab}, a member of the
Opera security team forwarded their response to us via email stating that: ``Opera on iOS giving background tabs access to the events does seem like an unwanted bug''. 

\textbf{Safari.} We reported to Apple about Safari allowing JavaScript access to sensor data within an iframe and also when the phone is locked. The Apple security team acknowledged the problem via email stating that: ``We have reviewed your paper and are working on the mitigations listed in the paper''. 

\textbf{W3C.} After we disclosed the identified problems to the W3C community, the community acknowledged the attack vectors introduced in this paper and stated that: ``This would be an issue to address for any future iterations on this document [i.e. W3C specification on mobile orientation and motion\cite{W3CMotion}]''. A security issue has been recorded to be taken into account by W3C in this regard\footnote{\texttt{github.com/w3c/deviceorientation/issues/13}}.
The community discussed this issue in their latest meeting and suggested to add a security section to the specification in response to the findings of our work\footnote{\texttt{w3.org/2015/10/26-geolocation-minutes.html\#item03}}. 

The industry feedback confirms that the currently unrestricted JavaScript access to certain sensor data within a web browser does present a serious security threat to the users. We appreciate the quick and constructive responses received from W3C and browser vendors, and hope the identified problems will be fixed in the near future.  
%==========================================
\section{Conclusion}
In this paper we introduced the first practical attack that was able to distinguish user touch actions as well as learning her PIN through JavaScript code embedded in a web page. 
We designed and implemented \emph{TouchSignatures}: a simple and effective JavaScript-based attack which when loaded within the browser was able to listen to the device orientation and motion sensor data streams and send the data back to a remote server for analysis. We demonstrated that \emph{TouchSignatures} was able to distinguish different user touch actions through a $k$-NN classifier, and PINs through ANN system, respectively. 
The results show that \emph{TouchSignatures} can classify user touch actions and identify her PIN digits with high success rates. 

Our results highlight major shortcomings in W3C standards, mobile operating systems, and browsers access control policy with respect to user security. 
As a countermeasure which strikes a balance between security and usability, we suggest that device orientation and motion data be treated similarly to GPS sensor data. Effective user notification and control mechanisms for access to such sensor data should be implemented both in mobile operating systems and in mobile browsers. The positive industry feedback confirms that serious damage could be caused exploiting the introduced attack vectors. As a matter of fact, some of the browser vendors such as Mozilla and Apple have already started working on the mitigations suggested in this paper. 

As future work, we would like to extend \emph{TouchSignatures} for other security purposes such as continues (implicit) authentication \cite{TouchMe,SilentSense,Progressiveauthentication:,mobileiden,useriden}.
%==========================================
\section{Acknowledgements}
We would like to thank the volunteers who contributed to the user studies of this project. We also thank several anonymous reviewers of this journal paper and its preliminary version at ASIACCS'15. We thank the W3C Geolocation Working Group and the mobile browser vendors including Mozilla, Apple, Google, and Opera for their quick responses and constructive communications. The last three authors are supported by ERC Starting Grant No.~306994.
%==========================================
\section{References}
%\bibliographystyle{abbrv}
%\bibliography{jsa-ref2}

\begin{thebibliography}{10}

\bibitem{W3CGPS}
{W3C Geolocation API Specification}.
\newblock http://dev.w3.org/geo/api/spec-source.html.

\bibitem{W3CLight}
{W3C Working Draft Document on Ambient Light Events}.
\newblock http://www.w3.org/TR/ambient-light/.

\bibitem{W3CMotion}
{W3C Working Draft Document on Device Orientation Event}.
\newblock http://www.w3.org/TR/orientation-event/.

\bibitem{W3CCamera}
{W3C Working Draft Document on Media Capture and Streams}.
\newblock http://w3c.github.io/mediacapture-main/getusermedia.html.

\bibitem{Tapprints2}
A.~J. Aviv, B.~Sapp, M.~Blaze, and J.~M. Smith.
\newblock Practicality of accelerometer side channels on smartphones.
\newblock In {\em Proceedings of the 28th Annual Computer Security Applications
  Conference}, pages 41--50. ACM, 2012.

\bibitem{SilentSense}
C.~Bo, L.~Zhang, X.-Y. Li, Q.~Huang, and Y.~Wang.
\newblock Silentsense: Silent user identification via touch and movement
  behavioral biometrics.
\newblock In {\em Proceedings of the 19th Annual International Conference on
  Mobile Computing and Networking}, MobiCom '13, pages 187--190, New York, NY,
  USA, 2013. ACM.

\bibitem{mobileiden}
H.~Bojinov, Y.~Michalevsky, G.~Nakibly, and D.~Boneh.
\newblock Mobile device identification via sensor fingerprinting.
\newblock {\em CoRR}, abs/1408.1416, 2014.

\bibitem{touchlogger}
L.~Cai and H.~Chen.
\newblock Touchlogger: Inferring keystrokes on touch screen from smartphone
  motion.
\newblock In {\em HotSec}, 2011.

\bibitem{Motionattack}
L.~Cai and H.~Chen.
\newblock On the practicality of motion based keystroke inference attack.
\newblock In S.~Katzenbeisser, E.~Weippl, L.~Camp, M.~Volkamer, M.~Reiter, and
  X.~Zhang, editors, {\em Trust and Trustworthy Computing}, volume 7344 of {\em
  Lecture Notes in Computer Science}, pages 273--290. Springer Berlin
  Heidelberg, 2012.

\bibitem{knn}
T.~Cover and P.~Hart.
\newblock Nearest neighbor pattern classification.
\newblock {\em Information Theory, IEEE Transactions on}, 13(1):21--27, 1967.

\bibitem{TouchMe}
A.~De~Luca, A.~Hang, F.~Brudy, C.~Lindner, and H.~Hussmann.
\newblock Touch me once and i know it's you!: Implicit authentication based on
  touch screen patterns.
\newblock In {\em Proceedings of the SIGCHI Conference on Human Factors in
  Computing Systems}, CHI '12, pages 987--996, New York, NY, USA, 2012. ACM.

\bibitem{Audio:Light}
T.~Halevi, D.~Ma, N.~Saxena, and T.~Xiang.
\newblock Secure proximity detection for {NFC} devices based on ambient sensor
  data.
\newblock In {\em Computer Security--ESORICS 2012}, pages 379--396. Springer,
  2012.

\bibitem{Tap-Wave-Rub}
H.~Li, D.~Ma, N.~Saxena, B.~Shrestha, and Y.~Zhu.
\newblock Tap-wave-rub: Lightweight malware prevention for smartphones using
  intuitive human gestures.
\newblock In {\em Proceedings of the Sixth ACM Conference on Security and
  Privacy in Wireless and Mobile Networks}, WiSec '13, pages 25--30, New York,
  NY, USA, 2013. ACM.

\bibitem{Speech:Gyr}
Y.~Michalevsky, D.~Boneh, and G.~Nakibly.
\newblock Gyrophone: Recognizing speech from gyroscope signals.
\newblock In {\em Proc. 23rd USENIX Security Symposium}, 2014.

\bibitem{Tapprints}
E.~Miluzzo, A.~Varshavsky, S.~Balakrishnan, and R.~R. Choudhury.
\newblock Tapprints: your finger taps have fingerprints.
\newblock In {\em Proceedings of the 10th international conference on Mobile
  systems, applications, and services}, pages 323--336. ACM, 2012.

\bibitem{ANNbook}
M.~F. Moller.
\newblock A scaled conjugate gradient algorithm for fast supervised learning.
\newblock {\em Neural Networks}, 6(4):525 -- 533, 1993.

\bibitem{KeyMic}
S.~Narain, A.~Sanatinia, and G.~Noubir.
\newblock Single-stroke language-agnostic keylogging using stereo-microphones
  and domain specific machine learning.
\newblock In {\em Proceedings of the 2014 ACM Conference on Security and
  Privacy in Wireless; Mobile Networks}, WiSec '14, pages 201--212, New York,
  NY, USA, 2014. ACM.

\bibitem{accessory}
E.~Owusu, J.~Han, S.~Das, A.~Perrig, and J.~Zhang.
\newblock Accessory: password inference using accelerometers on smartphones.
\newblock In {\em Proceedings of the Twelfth Workshop on Mobile Computing
  Systems \& Applications}, page~9. ACM, 2012.

\bibitem{Progressiveauthentication:}
O.~Riva, C.~Qin, K.~Strauss, and D.~Lymberopoulos.
\newblock Progressive authentication: deciding when to authenticate on mobile
  phones.
\newblock In {\em In Proceedings of 21st USENIX Security Symposium}, 2012.

\bibitem{All:Sensors}
B.~Shrestha, N.~Saxena, H.~T.~T. Truong, and N.~Asokan.
\newblock Drone to the rescue: Relay-resilient authentication using ambient
  multi-sensing.
\newblock In {\em Proc. Eighteenth International Conference on Financial
  Cryptography and Data Security}, 2014.

\bibitem{PINCamera}
L.~Simon and R.~Anderson.
\newblock Pin skimmer: Inferring pins through the camera and microphone.
\newblock In {\em Proceedings of the Third ACM Workshop on Security and Privacy
  in Smartphones \& Mobile Devices}, SPSM '13, pages 67--78, New York, NY, USA,
  2013. ACM.

\bibitem{SkimLight}
R.~Spreitzer.
\newblock Pin skimming: Exploiting the ambient-light sensor in mobile devices.
\newblock In {\em Proceedings of the 4th ACM Workshop on Security and Privacy
  in Smartphones \& Mobile Devices}, SPSM '14, pages 51--62, New York, NY, USA,
  2014. ACM.

\bibitem{useriden}
M.~Velten, P.~Schneider, S.~Wessel, and C.~Eckert.
\newblock User identity verification based on touchscreen interaction analysis
  in web contexts.
\newblock 9065:268--282, 2015.

\bibitem{taplogger}
Z.~Xu, K.~Bai, and S.~Zhu.
\newblock Taplogger: Inferring user inputs on smartphone touchscreens using
  on-board motion sensors.
\newblock In {\em Proceedings of the fifth ACM conference on Security and
  Privacy in Wireless and Mobile Networks}, pages 113--124. ACM, 2012.

\end{thebibliography}

%==========================================
\appendix
%==========================================
\section{Popular Browsers}
\label{appa}
\begin{table}[t] 
	\centering 
	\begin{tabular}{l l r}   
		\hline
		Name & Version & \#Downloads\ \ \ \\
		\hline
		Chrome & 40.0.2214.89 & 500,000,000+\\
		Opera Mini Fast Browser & 7.6.40234& 100,000,000+\\
		Opera browser for Android & 20.0.1656.87080& 50,000,000+\\
		Firefox & 34.0.1 & 50,000,000+\\
		Dolphin & 11.3.4& 50,000,000+\\
		UC Browser for Android & 10.1.0.527& 50,000,000+\\
		UC Browser Mini for Android & 9.7.0.520& 10,000,000+\\
		UC Browser HD & 3.4.3.532& 10,000,000+\\
		Baidu Browser (fast and secure) & 4.6.0.6& 10,000,000+\\
		CM Browser Fast \& Secure &5.1.44 & 10,000,000+\\
		Mobile Classic (Opera-based) & N/A & 10,000,000+\\
		Photon Flash Player \& Browser & 4.8 & 10,000,000+\\
		Maxthon Browser Fast & 4.3.7.2000& 5,000,000+\\
		Boat Browser for Android & 8.2.1& 5,000,000+\\ 
		Next Browser for Android & 1.17& 5,000,000+\\    
		Yandex.Browser & 14.12& 5,000,000+\\
		\hline 
		\end{tabular}
	\caption{Popular Android web browsers with full capabilities. Browsers with limited capabilities that do not support multi-tab browsing are excluded. The numbers of downloads were obtained from the Google Play Store, Jan 2015.}
	\label{tbl:dl}
\end{table}
%==========================================
We tested several browsers including three major browsers on Android: Chrome, Firefox, and Opera, and three major browsers on iOS: Safari, Chrome, and Opera. 
Other Android browsers were also included in the study due to their high download counts on the Google Play Store. 
The full list of tested Android browsers and their download counts can be seen in Table~\ref{tbl:dl}. 
There are a number of browsers with high numbers of downloads but limited capabilities, e.g., specialised search engine browsers or email-based browsers. Since these browsers do not support features such as multi-tab browsing, they are excluded from our study. 
The iOS App Store does not report the number of downloads, hence we used a combination of user ratings, iTunes Charts, and checking the availability of the listed Android browsers on iOS to discover and select a list of popular browsers on iOS. 
On both platforms, we only considered browsers that are available free of charge from the official app stores. 
%==========================================
\section{JavaScript code to access motion and orientation data}
\label{appb}
\lstset{
	language=C,
	basicstyle=\footnotesize\sffamily,
	frame=tblr,
	backgroundcolor=\color{yellow!20},
	numbers=left,
	xleftmargin=5.0ex,
	numberstyle=\tiny,
	numbersep=4pt,
	stepnumber=2,
	showstringspaces=false,
	keywordstyle=\color{blue}\bfseries
}

\lstset{emph={%  Adjust any special keywords
		printf
	},emphstyle={\color{red}\bfseries}
}
%==========================================
Our JavaScript code, used in different parts of the paper, sends the orientation and motion sensor data of the mobile device, if accessible through the testing browser, to our NoSQL database on mongolab.com. When the event listener fires, it establishes a socket (by using Socket.IO) between the client and the server and continuously transmits the sensor data to the database. 
%==========================================
\begin{lstlisting}
function socketInit(){
//initial settings
socket= io.connect();
socket.on('connected', function(){
if (window.DeviceOrientationEvent){
window.addEventListener('deviceorientation', function(event){
var gamma= event.gamma; 
var beta= event.beta; 
var alpha= event.alpha; 
socket.emit('OX', gamma); 
socket.emit('OY', beta); 
socket.emit('OZ', alpha); }); }
if (window.DeviceMotionEvent){
window.addEventListener('devicemotion', function(event){
var acceleration= event.acceleration;
var gacc= event.accelerationIncludingGravity;
var rotationRate= event.rotationRate;
var interval= event.interval;
var ax= acceleration.x; 
var ay= acceleration.y;  
var az= acceleration.z;			
var ralpha= rotationRate.alpha;
var rbeta= rotationRate.beta;
var rgama= rotationRate.gamma;
var gx= gacc.x;  var gy= gacc.y;  var gz= gacc.z;				
socket.emit('MX', ax); 
socket.emit('MY', ay); 
socket.emit('MZ', az);
socket.emit('rAlpha', ralpha); 
socket.emit('rBeta', rbeta); 
socket.emit('rGama', rgama);
socket.emit('MGX', gx); 
socket.emit('MGY', gy); 
socket.emit('MGZ', gz);
socket.emit('interval', interval); }); }
socket.on('disconnect', function(){		
alert("Disconnected!"); }); }
\end{lstlisting}
%==========================================
\end{document}